\documentclass[prb ,aps,twocolumn,groupaddress]{revtex4-1}
\usepackage{titlesec}
\titleformat*{\section}{\bf\fontsize{12}{2}\selectfont}
\usepackage[latin1]{inputenc}
\usepackage{amsmath}
\usepackage{amsfonts}
\usepackage{bm}
\usepackage{amssymb}
\usepackage{graphicx,bm}
\usepackage{mathrsfs}
\usepackage{hyperref}
\usepackage{esint}
\usepackage[caption=false]{subfig}

\newcommand{\bpm}{\begin{pmatrix}}
\newcommand{\lp}{\left(}

\newcommand{\lb}{\left[}

\newcommand{\rb}{\right]}

\newcommand{\rp}{\right)}

\newcommand{\epm}{\end{pmatrix}}

\newcommand{\ba}{\begin{eqnarray}}
\newcommand{\ban}{\begin{eqnarray}\nn}
\newcommand{\ea}{\end{eqnarray}}
\newcommand{\nn}{\nonumber}
\begin{document}
\author{Christian Boyd, Philip W. Phillips}
\affiliation{Department of Physics and Institute for Condensed Matter Theory,
	University of Illinois 1110 W. Green Street, Urbana, IL 61801, U.S.A}
\title{Single-parameter scaling in the magnetoresistance of optimally doped La$_{2-x}$Sr$_{x}$CuO$_4$}
\begin{abstract}
We show that the recent magnetoresistance data on thin-film La$_{2-x}$Sr$_{x}$CuO$_4$ (LSCO) in strong magnetic fields ($B$)\cite{Boebinger18} obeys a single-parameter scaling of the form MR$(B,T)=f(\mu_H(T)B)$, where $\mu_H^{-1}(T)\sim T^{\alpha}$ ($1\le\alpha\le2$), from $T=180$K until $T\sim20$K, at which point the single-parameter scaling breaks down.  The functional form of the MR is distinct from the simple quadratic-to-linear quadrature combination of temperature and magnetic field found in the optimally doped iron superconductor BaFe${}_2$(As${}_{1-x}$P${}_x$)${}_2$ \cite{Shekhter16}.  Further, low-temperature departure of the MR in LSCO from its high-temperature scaling law leads us to conclude that the MR curve collapse is not the result of quantum critical scaling.  We examine the classical effective medium theory (EMT) previously\cite{Patel18} used to obtain the quadrature resistivity dependence on field and temperature for metals with a $T$-linear zero-field resistivity.  It appears that this scaling form results only for a binary, random distribution of metallic components.  More generally, we find a low-temperature, high-field region where the resistivity is simultaneously $T$ and $B$ linear when multiple metallic components are present.  Our findings indicate that if mesoscopic disorder is relevant to the magnetoresistance in strange metal materials, the binary-distribution model which seems to be relevant to the iron pnictides is distinct from the more broad-continuous distributions relevant to the cuprates. Using the latter, we examine the applicability of classical effective medium theory to the MR in LSCO and compare calculated MR curves with the experimental data.
\end{abstract}
\maketitle

\section{Introduction}

A new ingredient uncovered recently in the study of strange metal physics in strongly correlated electron systems is the linear in B growth of the magnetoresistance (MR)\cite{Shekhter16,Hussey19,Boebinger18}.  First observed\cite{Shekhter16} in the iron superconductor BaFe$_2$(As$_{1-x}$P$_x$)$_2$ is a scaling collapse of the resistivity ($\rho$) data with a quadrature of the form $[\rho(T,B)-\rho_0]=\sqrt{\lp\alpha k_B T\rp^2+\lp\gamma\mu_B B\rp^2}$ with dimensionless $\alpha,\gamma$ and $T$ and $B$ the temperature and magnetic field, respectively.   This scaling form is intriguing as it suggests a linear in $B$ scattering rate in addition to the Planckian rate\cite{Zaanen} characterizing strange metal physics.  Since then, the iron chalcogenide FeSe$_{1-x}$S$_{x}$\footnote{In the case of FeSe$_{1-x}$S$_{x}$, a subtraction procedure\cite{Hussey19} was used to remove a quadratic-in-field MR contribution before the quadrature temperature/field scaling was seen.  The quadratic MR has been attributed to the standard MR of metals due to scattering along Fermi surface orbits.} was also observed to exhibit linear MR at strong fields\cite{Hussey19} near a nematic critical point and similar behavior is seen in non-superconducting Ba(Fe$_{1/3}$Co$_{1/3}$Ni$_{1/3}$)$_2$As$_2$ near its magnetic critical point\cite{Nakajima19}. 

The magnetoresistance in La$_{2-x}$Sr$_{x}$CuO$_4$ (LSCO) tells a different story.   An experimental collaboration \cite{Boebinger18} observed large, unsaturating quadratic-to-linear magnetoresistance (MR) near optimal doping which does not obey the quadrature scaling. At low temperatures, a region of {\it simultaneous} $T,B$ linearity exists in strong fields ($B\sim 50-80$T).

\begin{figure}[h!]
	\centering
	\includegraphics[width=\linewidth]{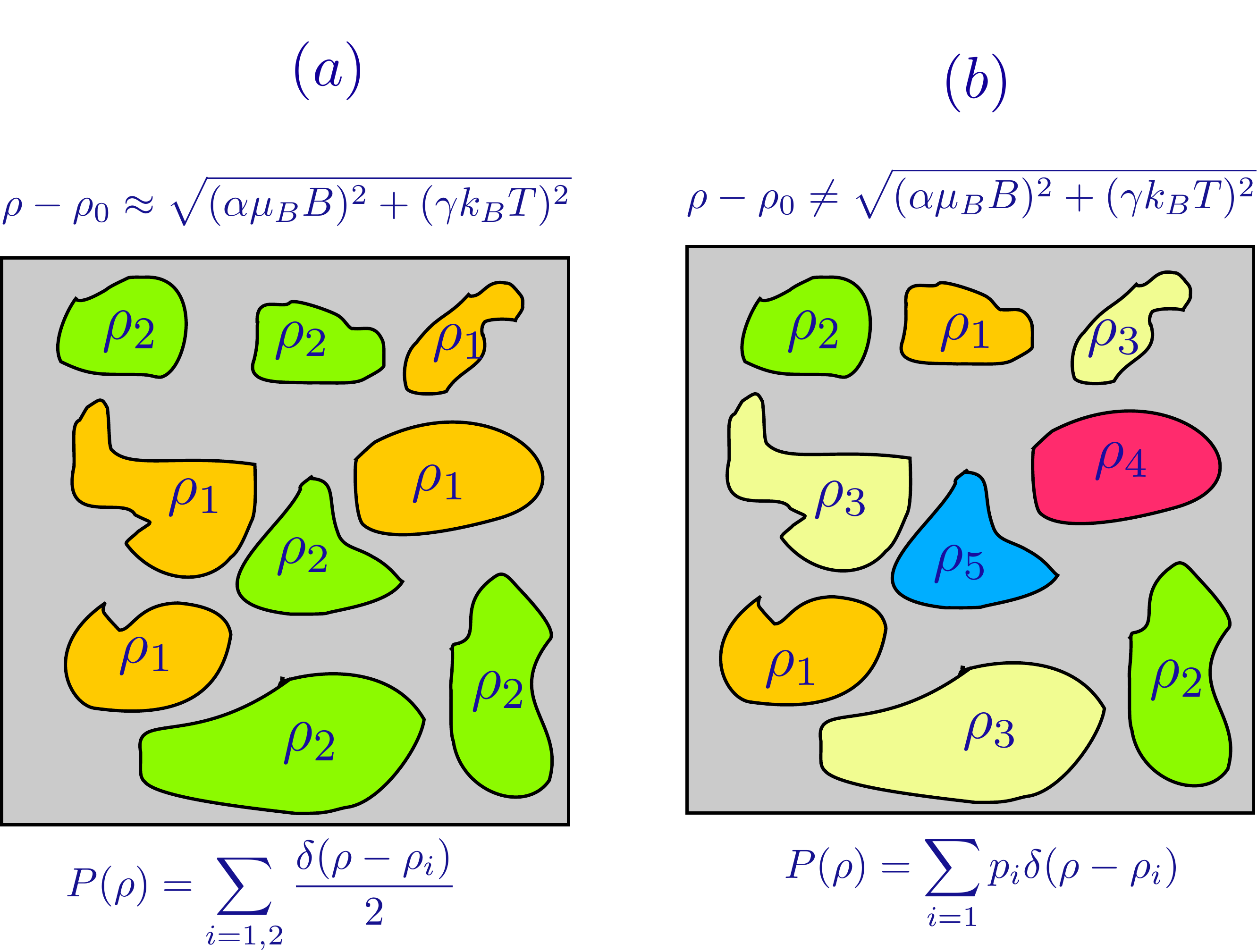}
	\caption{A visual summary of our investigation into effective medium theory applied to an inhomogeneous material where each region contains a $T$-linear resistivity ({\bf Note}: not to scale).  In systems other than those with two components (equally distributed), the MR is distinctly {\it not} given by the quadrature combination of field and temperature scales $\rho(T,B)-\rho_0=\sqrt{\lp\alpha k_B T\rp^2+\lp\gamma\mu_B B\rp^2}$.}
	\label{figemt}
\end{figure}

In trying to reconcile the temperature dependence of LSCO with the iron quadratic-to-linear MR curves, we uncover a new single-parameter scaling in the LSCO MR data.  By examining the effective medium theories that have been used to understand B-linear resistivity in inhomogeneous materials, we show that while such programs can yield quadrature temperature/field scaling, such a combination {\it only} applies to the case of an equally-distributed  two-component system as depicted in Fig. (\ref{figemt}a).  In the general case of multiple metallic constituents, Fig. (\ref{figemt}b), or in the continuum limit, we find temperature scaling more closely resembling the LSCO experiment -- i.e., linear in both temperature and applied magnetic field at low temperature and strong fields.  We close by examining the validity of applying a classical theory to the MR of LSCO and discuss possible issues.

\section{New Single-parameter Scaling in the Magnetoresistance}

\begin{figure}[h!]
	\centering
	\subfloat[\label{sfig:1a}]{
	\includegraphics[width=\linewidth]{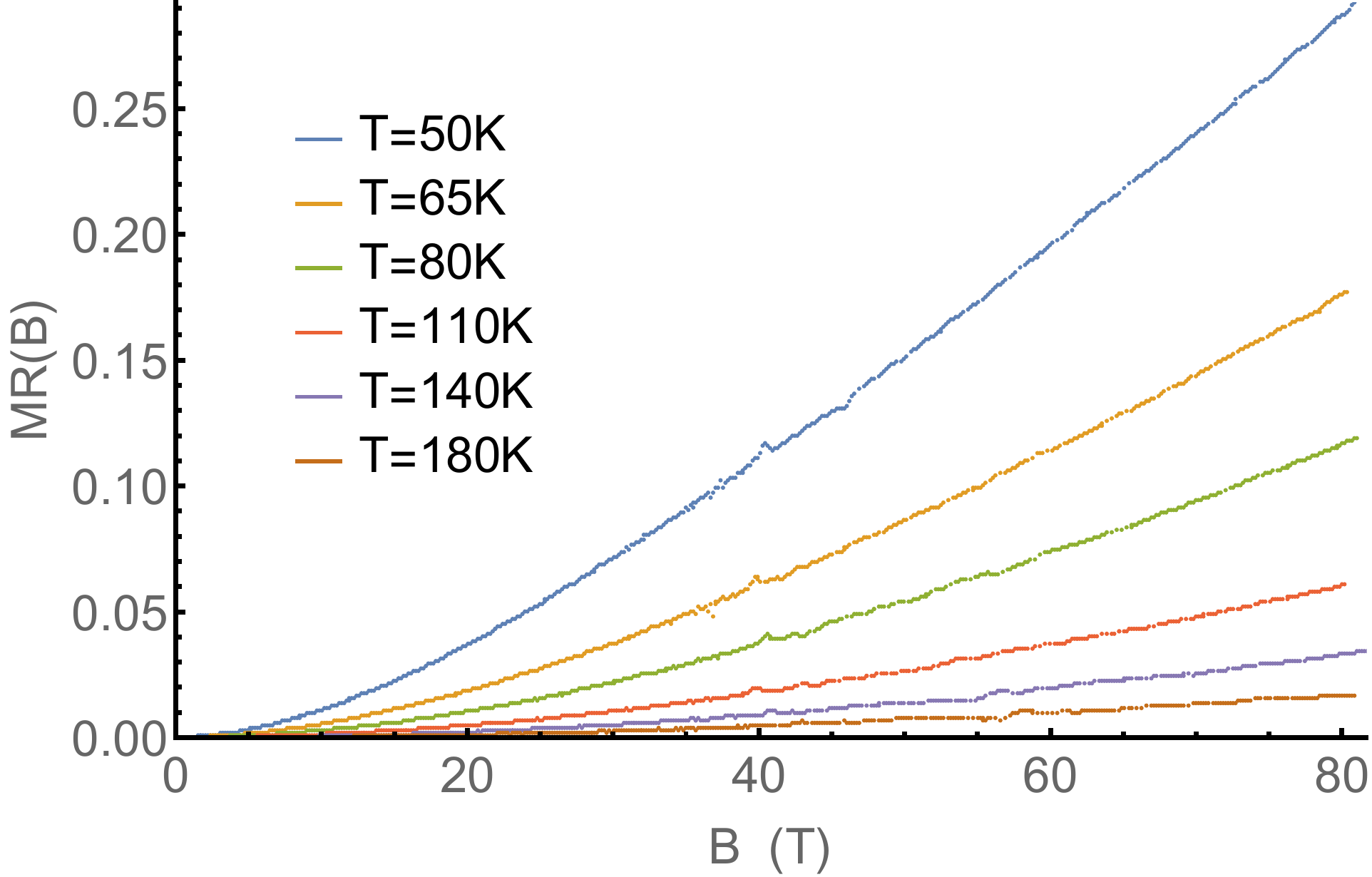}
	}

	\subfloat[\label{sfig:1b}]{
		\includegraphics[width=\linewidth]{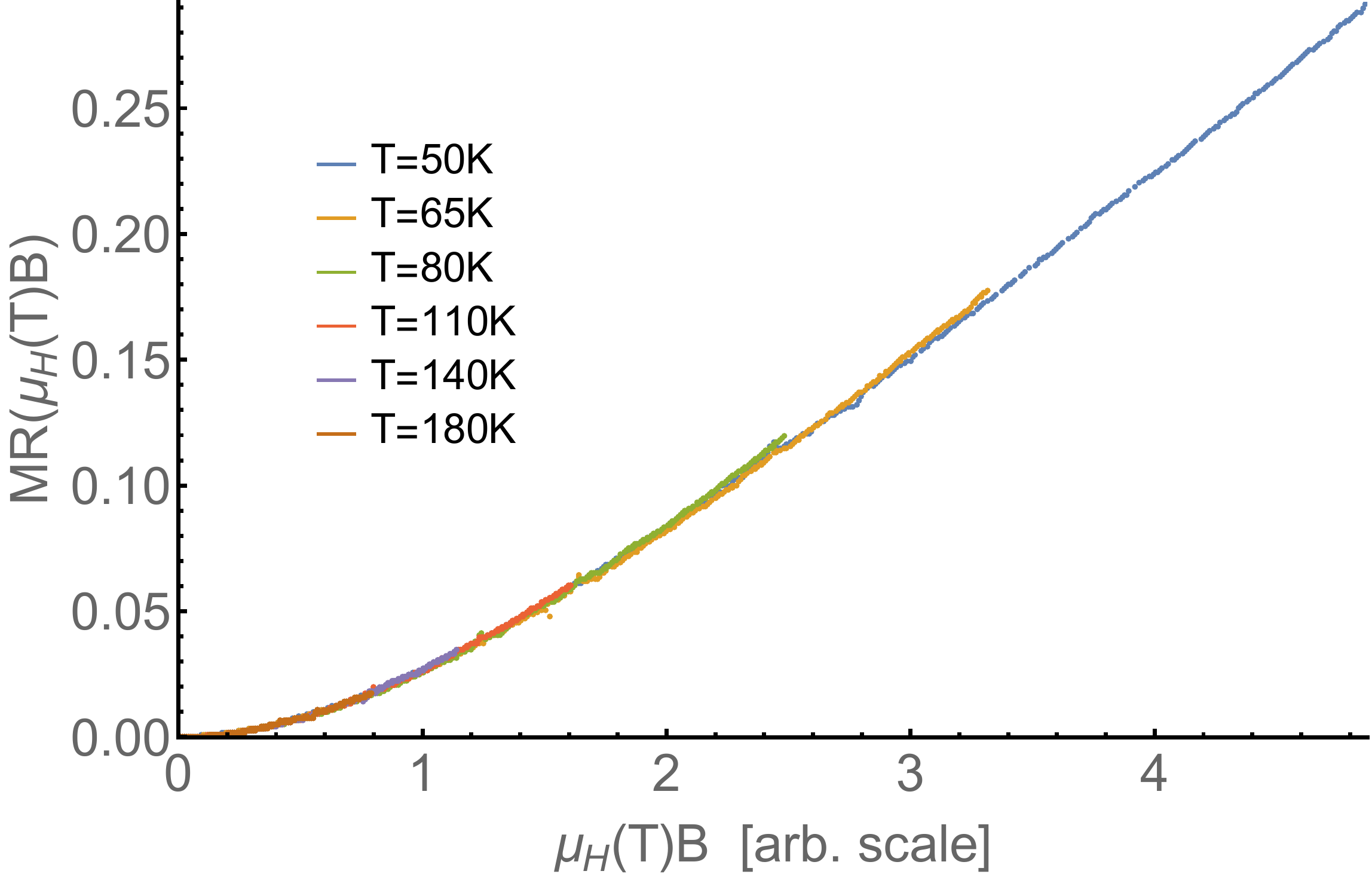}
	}
	\caption{{\bf (a)}  The unscaled magnetoresistance curves for LSCO at doping $p=0.19$ using data available online (See Supplementary Materials\cite{Boebinger18}).  {\bf (b)}  Curve collapse of the magnetoresistance data at temperatures above $T_c$ by rescaling the magnetic field with a temperature-dependent factor $\mu_H(T)$.  Plotted are the raw data points rescaled by the temperature-dependent scale $\mu_H(T)$ [Fig. 2(b)].  The overall magnitude of $\mu_H(T)$ is undetermined.}
	\label{fig1}
\end{figure}

\begin{figure}[h!]
	\centering
	\subfloat[\label{sfig:2a}]{
		\includegraphics[width=\linewidth]{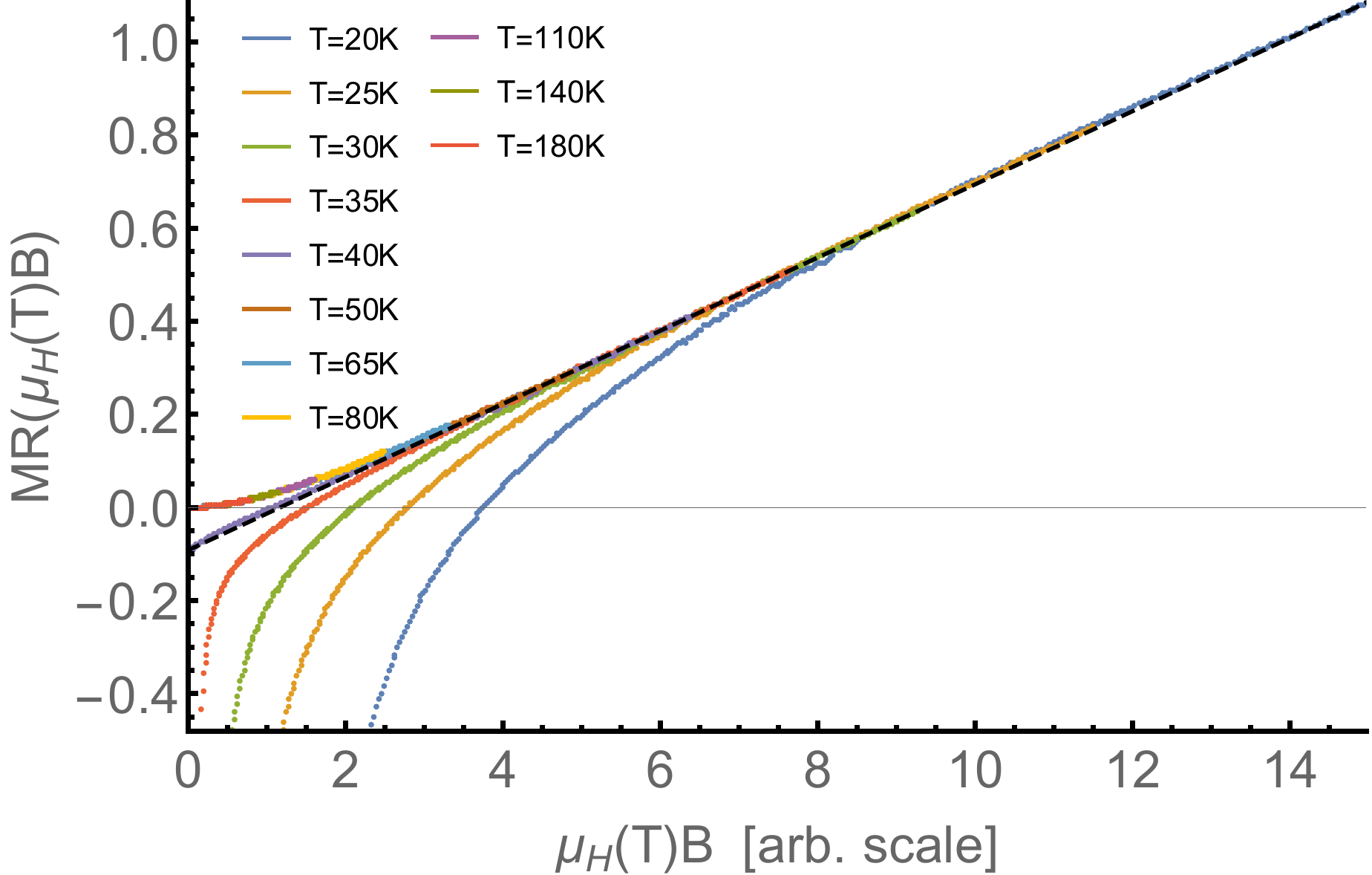}
	}
	
	\subfloat[\label{sfig:2b}]{
		\includegraphics[width=\linewidth]{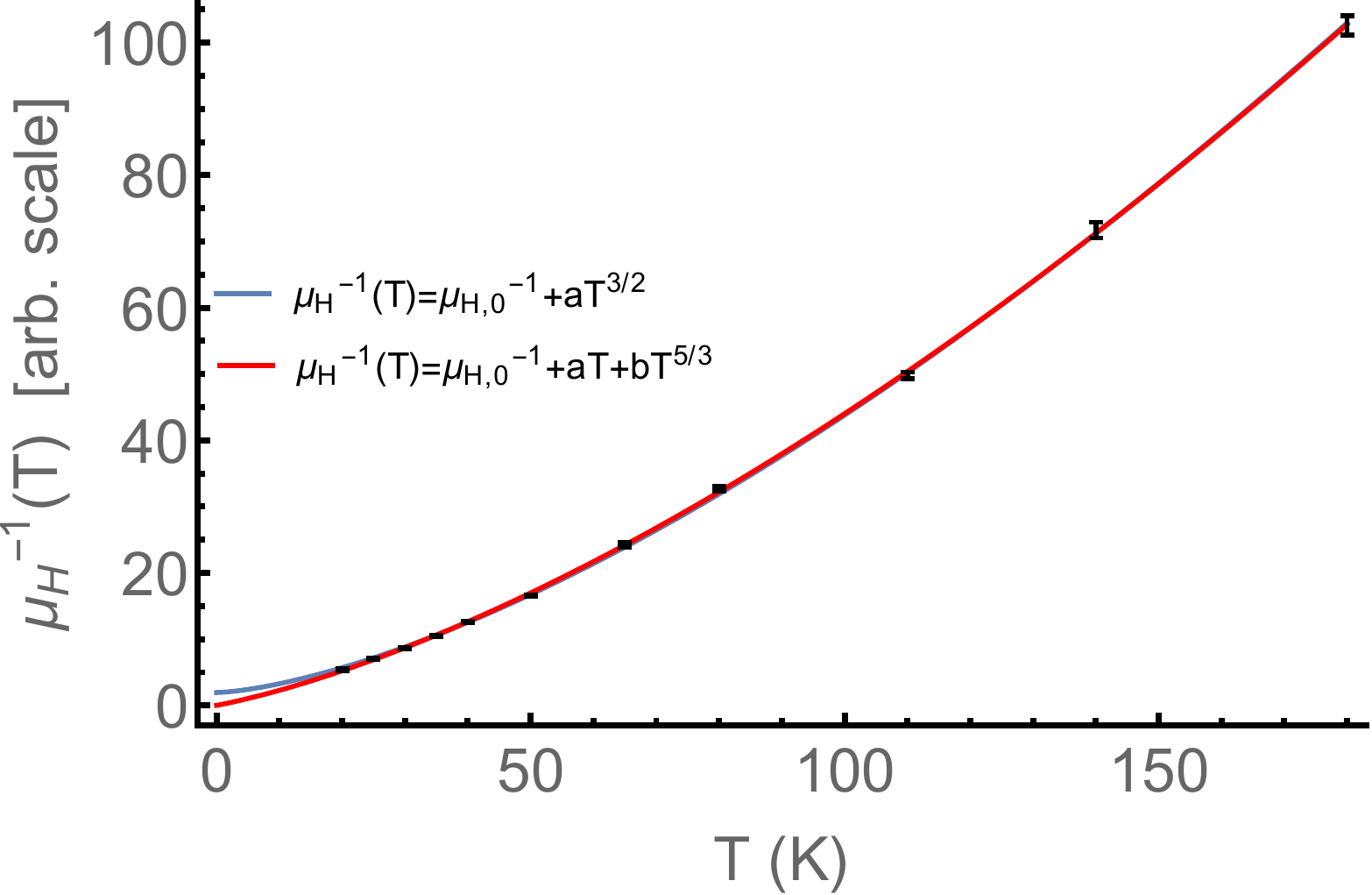}
	}
	\caption{{\bf(a)} The collapse shown in Fig. \ref{sfig:1b} extended below $T_C$ using the experimental fit to the zero-field resistivity. {\bf (b)}  The temperature dependent magnetic field scale $\mu_H(T)$.  The points in $\mu_H(T)$ are calculated by inverting the scaling used to obtain curve collapse in (a).  Error bars represent the accuracy of the curve collapse fit defined by the minimum of the mean-squared error, not experimental error.  Filled curves are two possible fits to the calculated points.}
	\label{fig2}
\end{figure}

There exists a great deal of data (see Supplementary Materials\cite{Boebinger18}) on hole-doped LSCO samples ($p=0.161-0.19$) in the presence of strong magnetic fields.  Analyzed in terms of the magnetoresistance MR$(T,B) :=\lb\rho(T,B)-\rho(T,B=0)\rb/\rho(T,B=0)$ across the measured temperatures $T_C<50$K $<T<180$K, the experimental data produce several temperature-dependent curves (Fig. \ref{sfig:1a}).  Hidden within this data is a single temperature/magnetic field combination $\mu_H(T) B$ which collapses all of these different temperature curves onto a single MR curve (Fig. \ref{sfig:1b}).  In the case of the optimally doped iron superconductor\cite{Shekhter16} BaFe$_2$(As$_{1-x}$P$_x$)$_2$, iron chalcogenide\cite{Hussey19} FeSe$_{1-x}$S$_{x}$, and\cite{Nakajima19} Ba(Fe$_{1/3}$Co$_{1/3}$Ni$_{1/3}$)$_2$As$_2$, the MR was found to collapse onto a simple function MR$(B)\sim [\sqrt{1+\# B^2/T^2}-1]=f(B/T)$ in terms of a single parameter $(B/T)$ related to the zero-field scattering rate $\rho(T,B=0)\propto T$.  In contrast, the temperature dependence of the scaling parameter $\mu_H^{-1}(T)$ within the LSCO data (Fig. \ref{sfig:2b}) appears distinct from scattering rates\cite{Ando97,Batlogg94} inferred from the temperature dependence in the zero field resistivity $\rho(T)\sim \rho_0+\alpha T$ or the high temperature Hall angle $\cot\theta_H(T)\sim T^2$.  Further, if the experimental fit to the resistivity $\rho(T) = \rho_0 + \alpha T$ is used below $T_c$, the MR collapses onto a single quadratic-to-linear curve from $T=180$K all the way down to $T\sim 20$K (Fig. \ref{sfig:2a}) -- along which, the form of $\mu_H(T)$ does not appear altered from its high temperature scaling.

\begin{figure}[h!]
	\centering
	\subfloat[\label{sfig:3a}]{
		\includegraphics[width=\linewidth]{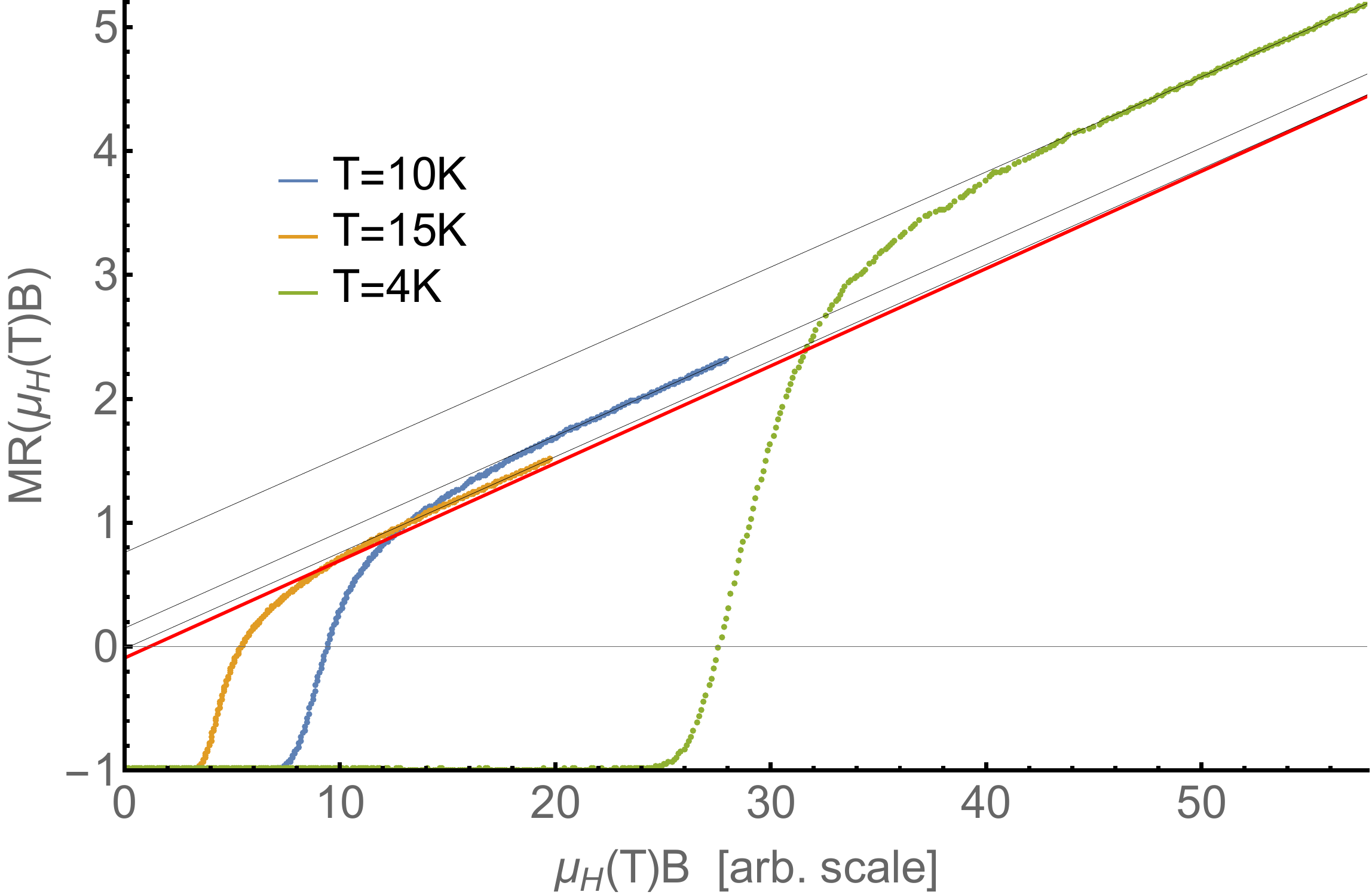}
	}
	
	\subfloat[\label{sfig:3b}]{
		\includegraphics[width=.49\linewidth]{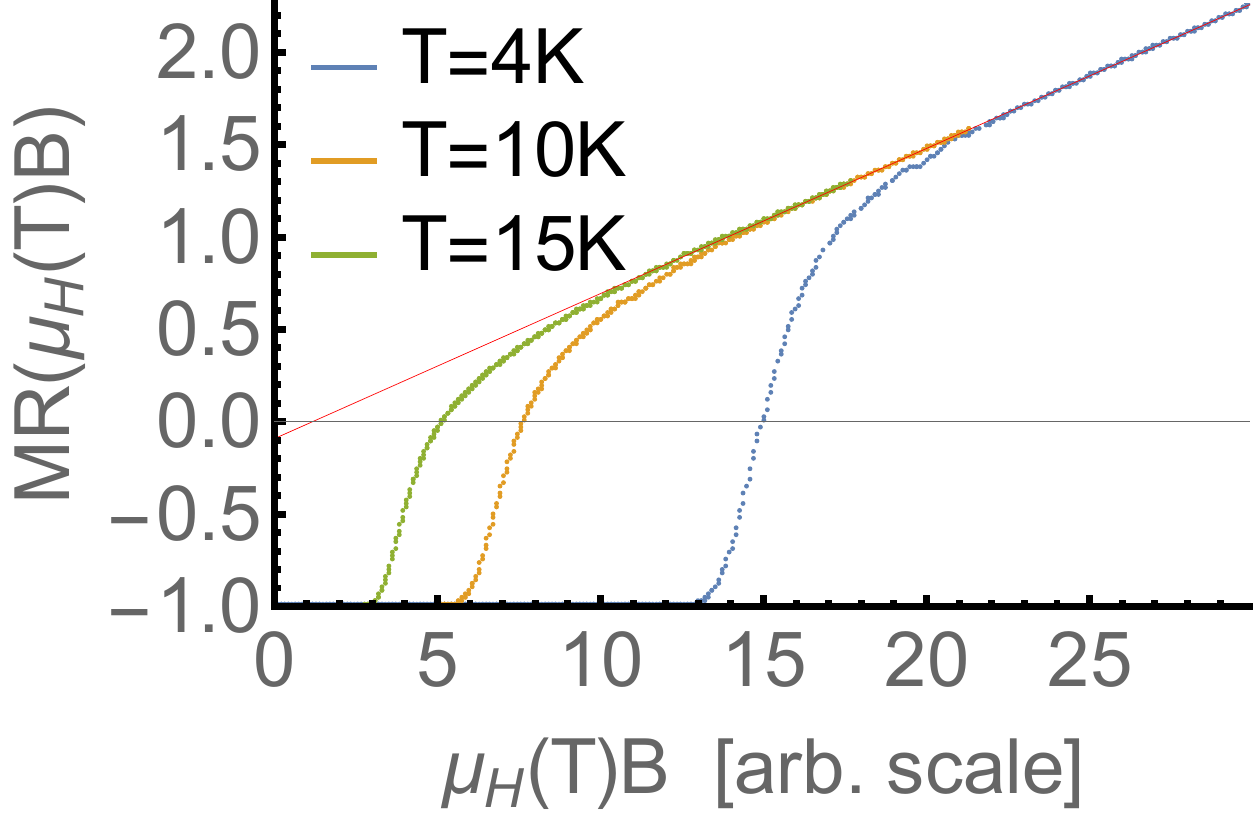}
	}
	\subfloat[\label{sfig:3c}]{
		\includegraphics[width=.49\linewidth]{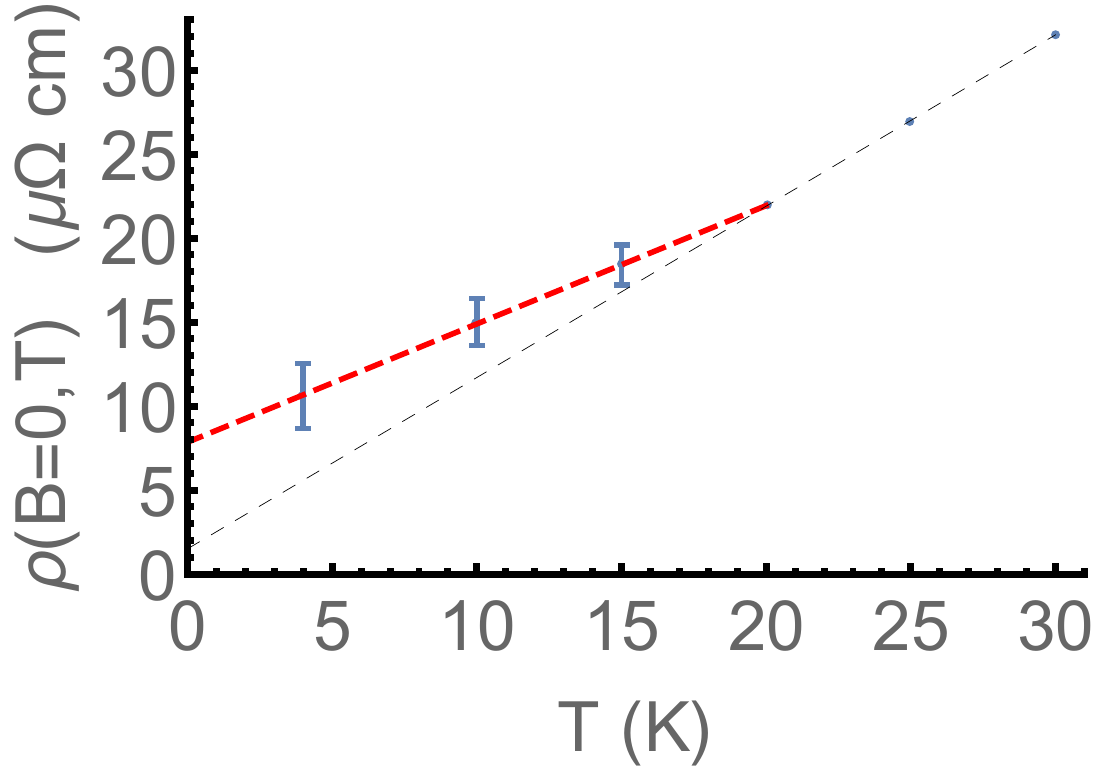}
	}
	\caption{{\bf (a)}  The magnetoresistance at $T\le15$K scaled to have the slope of the previous temperature fits [Fig. 2(c)] (with the same accuracy that determined $\mu_H(T)$). Black lines are high-field linear fits and the red line is the linear extrapolation of $T=50$K data.  Even if the residual resistivity $\rho_0$ is tuned within experimental uncertainty $\rho_0\le4.5\mu\Omega$cm, these 3 curves do not collapse for any $\mu_H(T)$.  {\bf (b)}  The result of enforcing single-parameter scaling by altering the zero-field resistivity $\rho(B=0,T)$ below $T\sim20$K.  The red line is the dashed line demonstrating curve collapse in Fig. \ref{sfig:2b}.  {\bf (c)}  The low-temperature resistivity inferred from the curve collapse in (b).  Error bars represent propagated error from uncertainty in $\mu_H(T)$ through the definition of the magnetoresistance MR$(B,T)$, not experimental error.  The red curve represents a linear fit to the form of the low-temperature resistivity necessary for curve collapse.}
	\label{fig3}
\end{figure}

Below $T\sim20$K, the high-field (50T-80T) MR is also linear in field but is no longer an extension of the curve extrapolated from higher temperatures (Fig. \ref{fig3}).  Interestingly, $T\le20$K is the regime in which the field derivative of the resistivity, $\beta:=\partial\rho(T,B)/\partial B$, appears to suddenly become temperature-independent at high field \cite{Boebinger18}.  If the MR is forced to collapse at low temperatures, then the inferred zero-temperature resistivity $\rho(T,B=0)$ must change abruptly near $T\sim20$K (Fig. \ref{sfig:3c}), though it could remain linear, but at the same time does not appear to upturn as would be expected from a residual, intervening pseudogap order.  Additionally, the field scaling $\mu_H(T)$ below $T=20$K would need to be proportional to this zero-field resistivity $\mu_H^{-1}(T)\propto\rho(T,B=0)$ in order to be consistent with the saturation of  $\beta$ to a constant value -- a fact that can be seen by differentiating MR$(B,T)$ and using the single-parameter assumption MR$(B,T)=f(\mu_H(T)B)$ when $\beta=$const.:
\ba
\frac{\partial\text{MR}(B,T)}{\partial B} &=& \mu_H(T)\frac{df(x)}{dx}\Big|_{x=\mu_H(T)B}
\\
&=&\frac{\beta}{\rho(B=0,T)}
\\
\lp\frac{df(x)}{dx},\beta\,\,\text{const.}\rp&\implies& \mu_H(T)\propto\frac{1}{\rho(B=0,T)}\,\,\,.
\ea
Regardless of whether or not we accept the scaling assumption that MR$(B,T)=f(\mu_H(T) B)$, the high-temperature scaling is either broken or the low-temperature resistivity (and/or temperature scaling $\mu_H(T)$) is altered from its high-temperature form.  In the first case  where the scaling is broken, a quantum critical explanation for the high-temperature MR data is no longer viable, since the $T$-linear resistivity behavior persists.  Alternatively, if the scaling is enforced, it appears that the high-temperature strange metal is distinct from the $T\le20$K state in a strong magnetic field.  The disconnect between high-temperature and low-temperature MR observed in hole-doped LSCO is also seen in the electron-doped cuprate La$_{2-x}$Ce$_x$CuO$_4$ where the low-temperature MR exhibits linear-in-field scaling once superconductivity is suppressed \cite{Sarkar18} and, in that case, is the only regime attributed to quantum criticality.  That the LSCO MR data above $T\sim20$K is not smoothly connected to the low-temperature MR, and the resulting inconsistency with quantum critical scaling at higher temperatures, is one of our principal conclusions.

\begin{figure}[h!]
	\centering
	\subfloat[\label{sfig:4a}]{
		\includegraphics[width=.24\linewidth]{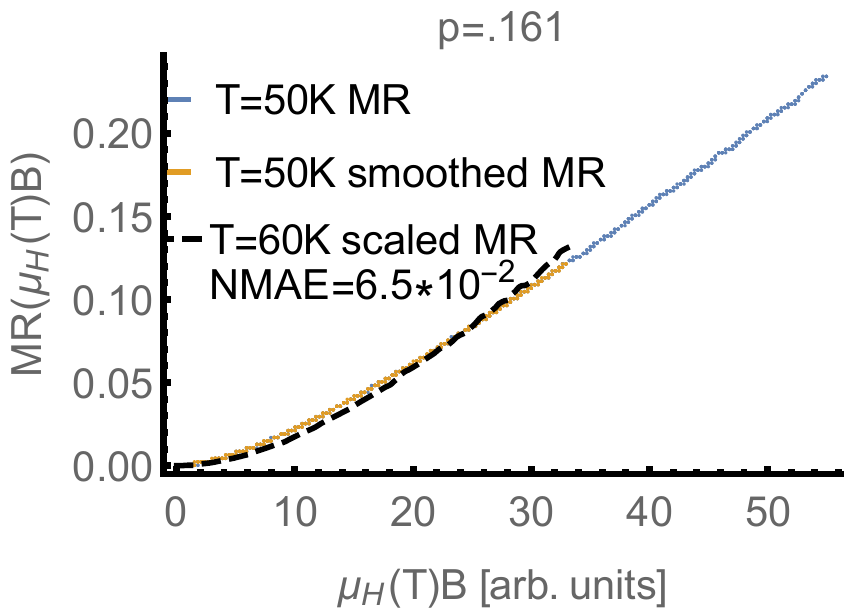}
	}
	\subfloat[\label{sfig:4b}]{
		\includegraphics[width=.24\linewidth]{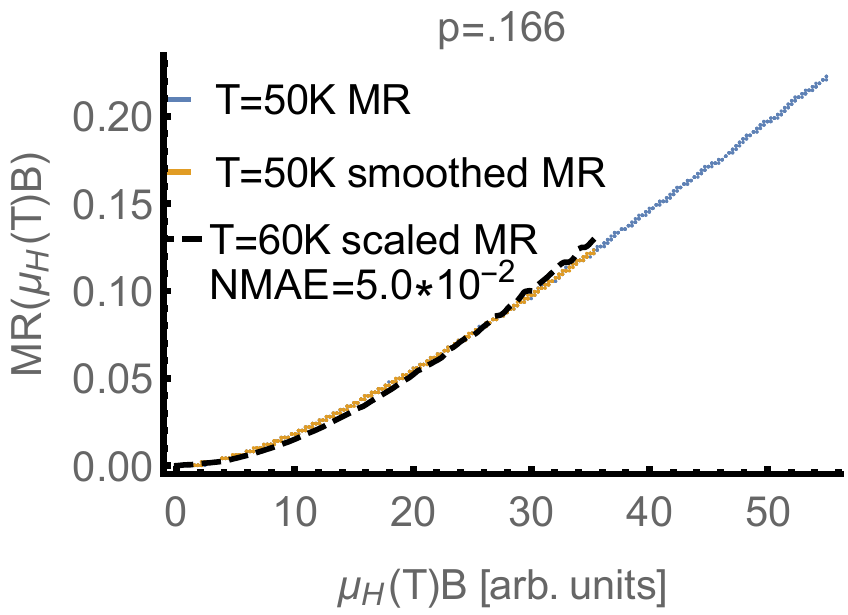}
	}
	\subfloat[\label{sfig:4c}]{
		\includegraphics[width=.24\linewidth]{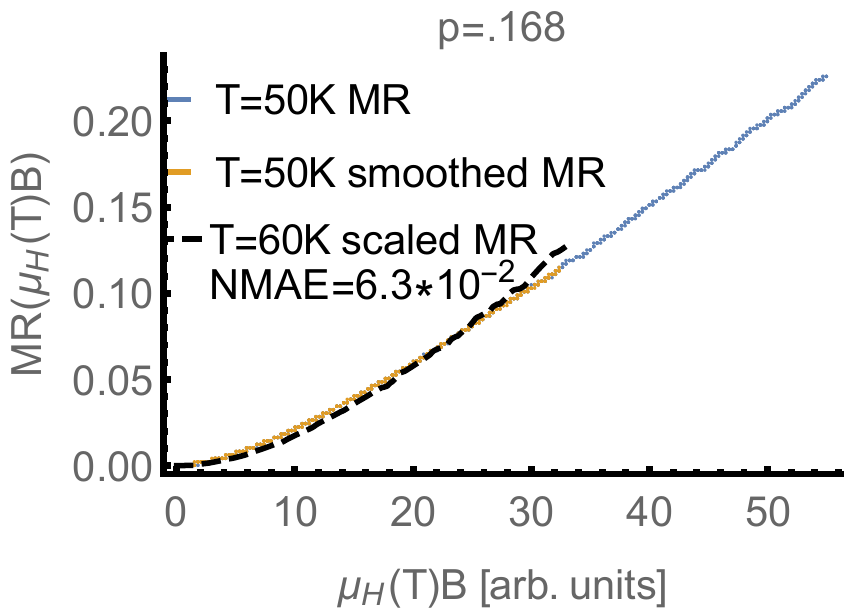}
	}
	\subfloat[\label{sfig:4d}]{
		\includegraphics[width=.24\linewidth]{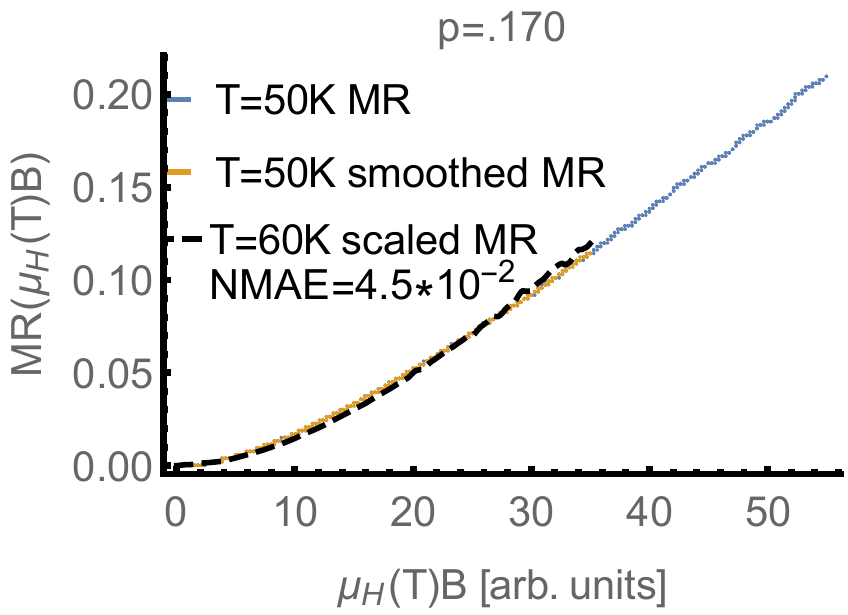}
	}
	
	\subfloat[\label{sfig:4e}]{
		\includegraphics[width=.24\linewidth]{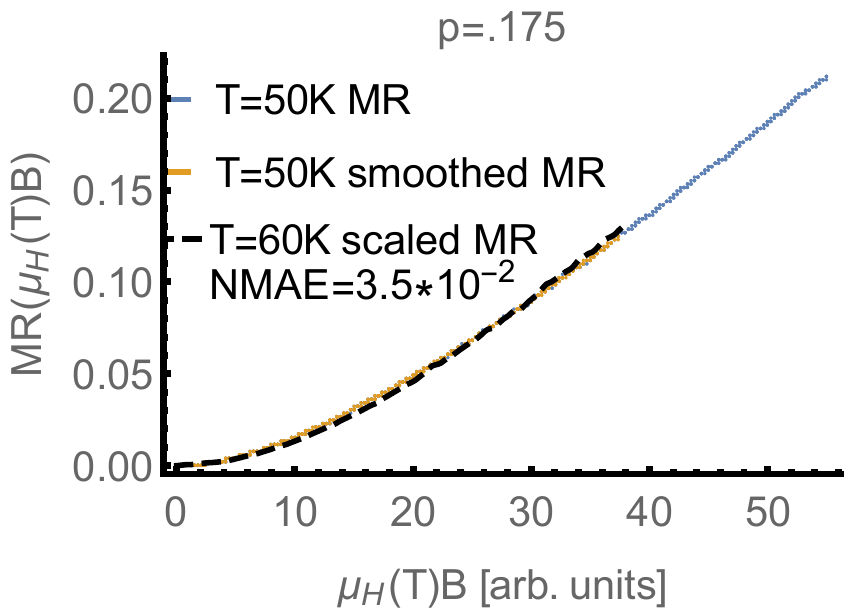}
	}
	\subfloat[\label{sfig:4f}]{
		\includegraphics[width=.24\linewidth]{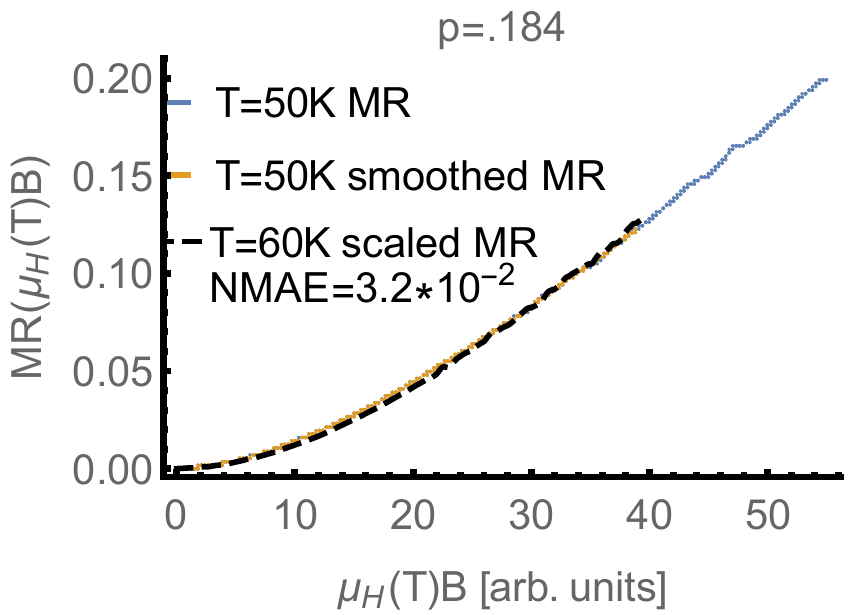}
	}
	\subfloat[\label{sfig:4g}]{
		\includegraphics[width=.24\linewidth]{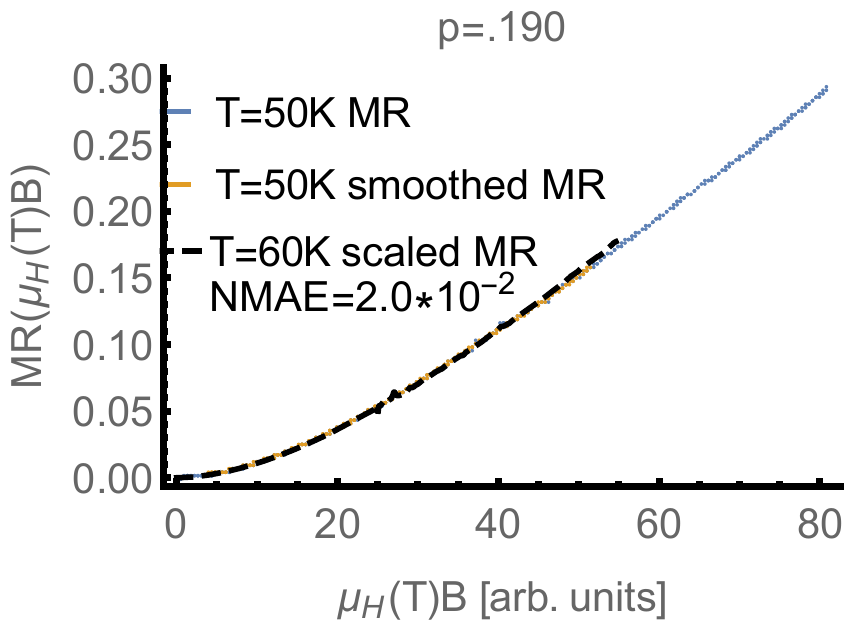}
	}
	
	\subfloat[\label{sfig:4h}]{
		\includegraphics[width=.49\linewidth]{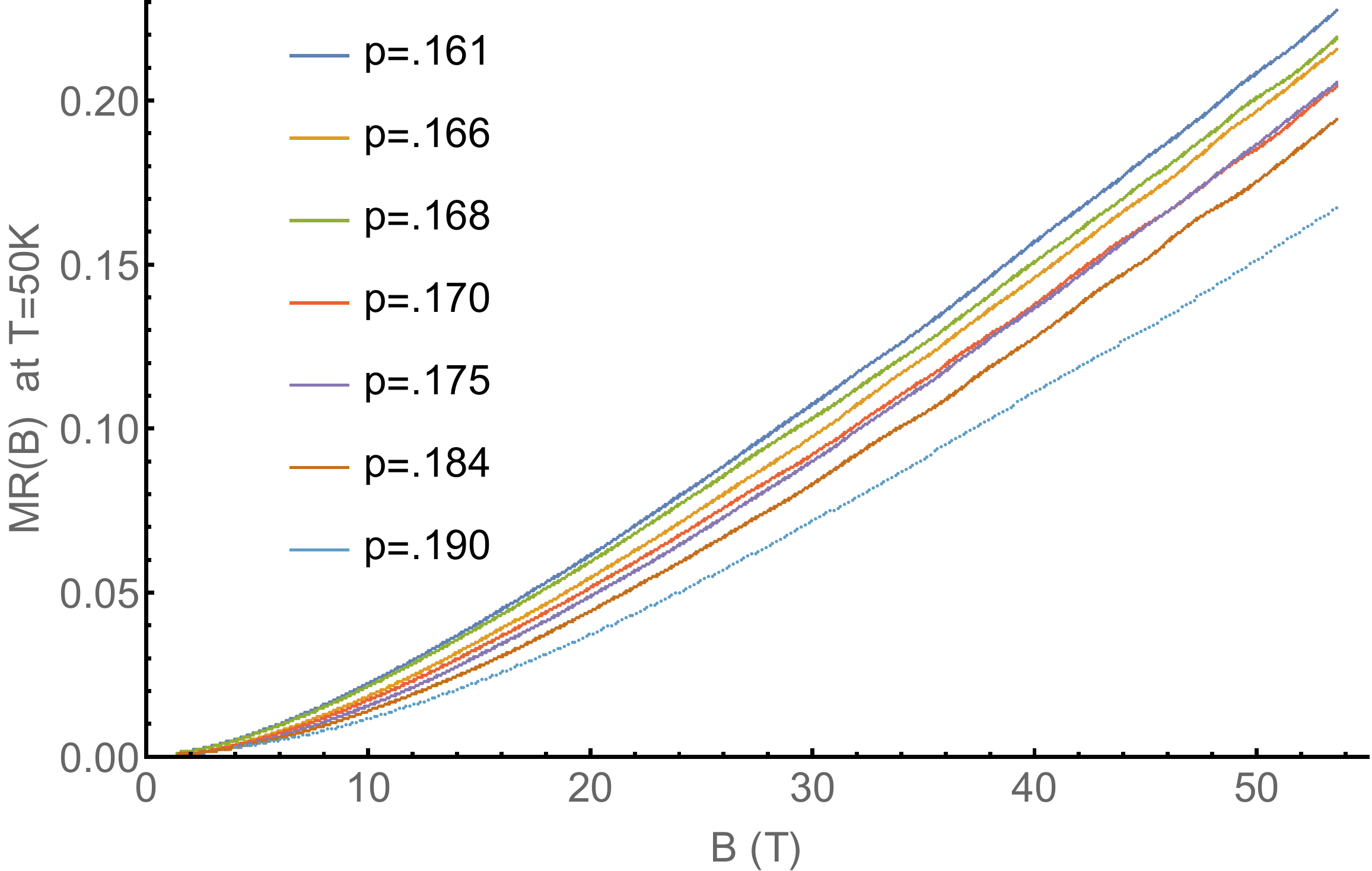}
	}
	\caption{{\bf (a)-(g)}  Collapse of the first two measured MR curves above $T_c$ in the doping range $p=.161-.190$, where $p=.190$ is the "optimally doped" sample analyzed in detail previously.  The Normalized Mean Absolute Error (NMAE) is compared to the smoothed $T=50$K MR data.  The "smoothing" used is a simple moving average over 50 data points.  The increase in error at $p=.168$ shown in (c)  relative to $p=.168$ may be due to the bump near the end of the data, similar to the end of the curve in the $p=.161$ shown in (a).  {\bf (h)}  Shown are the smoothed MR curves at $T=50$K across the doping range $p=.161-.190$.  The curves across this doping range are not continuously transforming into the MR at $p=.190$.  Notably, the $p=.168$ jumps up closer to the $p=.161$ curve, possibly related to the increase in error observed in (c).}
	\label{fig4}
\end{figure}

The resistivity data at high magnetic fields\cite{Boebinger18} was measured across several samples at different dopings $p=0.16-0.19$ -- all of which appear to demonstrate quadratic-to-linear magnetoresistance above $T_c$.  In contrast to the data taken at $p=0.19$, the curve collapse is not as absolute in the underdoped samples (Fig. \ref{fig4}).  The collapse appears to worsen below $p=.19$ in the more underdoped samples. By contrast, in the quantum critical iron compounds -- BaFe$_2$(As$_{1-x}$P$_{x}$)$_2$, Ba(Fe$_{1/3}$Co$_{1/3}$Ni$_{1/3}$)$_2$As$_2$, FeSe$_{1-x}$S$_{x}$ -- the simple magnetoresistance scaling MR$(B,T)\sim[\sqrt{1+\# B^2/T^2}-1]$ continues over an extended range of dopings -- e.g.\cite{Shekhter16}, in BaFe$_2$(As$_{1-x}$P$_{x}$)$_2$ from $p=.31$, optimal doping, to at least $p=.41$.  Further systematic MR data on the overdoped side of LSCO can clarify what behavior is unique to the doping $p=.19$.

\section{Effective medium theory and its continuum limits}

Previously, a theory collaboration \cite{Patel18} produced a quadrature combination of applied magnetic field and temperature in the resistivity $\rho(B,T)-\rho_0\propto\sqrt{\lp\alpha k_B T\rp^2+\lp\gamma\mu_B B\rp^2}$ by use of an effective medium theory (EMT).  The particular EMT\cite{Landauer78,Markel16} relevant to the current experiments is the extension to a tensor conductivity\cite{Stroud75} developed in the mid-'70s.  In this approach, one defines a  macroscopic conductivity $\sigma^E$ of a material made up of several constituents through the
\clearpage

\onecolumngrid

\twocolumngrid
\noindent
disorder-averaged local electric field $\bm E_0=\langle \bm E(x)\rangle$ and local current $\bm I(x)=\bm \sigma(x) \bm E(x)$ by $\bm \sigma^E \bm E_0 = \langle \bm I(x)\rangle$.  For local conducting patches with spherical (in 2D, circular) symmetry, $\sigma^E_{xx,xy}$ can be found self-consistently by replacing the external environment of a local conducting patch by its average and assuming the internal field is constant.  This approximation relies on the typical length scale of inhomogeneity being large enough to treat each region as having a well-defined local conductivity, e.g. if the domains are suitably larger than the mean free path, after which the conductivities are {\it classically} averaged using the aforementioned mean-field approximation.

The distribution of metallic components used to obtain the quadrature result \cite{Patel18} was chosen to be bi-valued.  At a fixed temperature (mobility), the MR of an inhomogeneous, two-component system is known to be described by the desired square root function\cite{Guttal05} when the two types of carriers are oppositely charged and each is present in exactly half the sample.  The results of EMT for same-sign charge carriers distributed equally throughout the system, as in the current case, was noted then\cite{Guttal05} as well.  One of the key points of inquiry in this work is whether or not this feature is robust to a more general multi-sourced distribution.  

In that same calculation, the scattering within each of the two types of metallic regions was assumed to be controlled by a mobility which captures strange-metallic behavior through $\mu\propto\mu_H\propto\tau\sim1/T$ where $\mu_H$ couples to the magnetic field.   In the standard case, $\mu_H B=\omega_C\tau$.  In Patel et al\cite{Patel18}, different slopes for the mobility scaling $\mu^{-1}(T)\propto T$ were used; however, the results are not affected by this change.

The magnetic field is assumed to affect the current only through the Lorentz force.  The steady state conditions, similar to that proposed in the Boltzmann equation in \cite{Harris95} are
\ba
\bm I &=& \sigma_0 \bm E + \mu_H \lp\bm I\times\bm B\rp
\\
\bm I &=& \bm \sigma(B)\cdot \bm E,
\ea
which lead to local, field-dependent conductivities of the form
\ba
\sigma_{xx}(x, B) &=& \frac{\sigma_0(x)}{1+\lp\mu_H B\rp^2}=\frac{\mu n(x)}{1+ \lp \mu_H B\rp^2}
\\ \label{Hall conductivity}
\sigma_{xy}(x,B) &=& \lp \mu_H B\rp\sigma_{xx}(x)=\frac{\mu n(x)\lp \mu_H B\rp}{1+\lp \mu_H B\rp^2}\,\,\,\,\,\,\,\,.
\ea
The zero-field conductivity is assumed to be locally Drude-like and isotropic $\sigma(x)=n(x) \mu$.  Hence, the full form of the macroscopic conductivity is given by the ansatz
\ba
\sigma^E(B) = 
\frac{\mu}{1+\lp\mu_H B\rp^2}\bpm
n^E_{xx} & \lp\mu_H B\rp n_{xy}^E
\\
-\lp \mu_H B\rp n_{xy}^E & n^E_{xx}
\epm\,\,\,\,\,.
\ea
The 2D tensorial extension of the Bruggeman-Landauer equation \cite{Stroud98} (or see\cite{Patel18,Guttal05}) results in the coupled effective medium equations
\ba
0 &=& \sum_{i=1,2}\frac{n_i^2-\lp n_{xx}^E\rp^2+\lp\mu_H B\rp^2\lp n_i-n_{xy}^E\rp^2}{\lp n_i+n_{xx}^E\rp^2+\lp\mu_H B\rp^2\lp n_i-n_{xy}^E\rp^2}
\\
0 &=& \sum_{i=1,2}\frac{n_i-n_{xy}^E}{\lp n_i+n_{xx}^E\rp^2+\lp\mu_H B\rp^2\lp n_i-n_{xy}^E\rp^2}\,\,\,\,,
\ea
where the $n_{1,2}$ refer to local carrier densities.  The exact solution is transparent in its dimensionless field strength $(\mu_H B)$ dependence,
\ba
n_{xx}^E &=&
\sqrt{n_1 n_2}\sqrt{1+\lp\frac{n_1-n_2}{n_1+n_2}\rp^2\lp\mu_H B\rp^2}
\\
n_{xy}^E &=& \frac{2n_1n_2}{\lp n_1+n_2\rp}\,\,\,.
\ea
Rescaling by the field/mobility factors to obtain the $\sigma^E(B)$, then inverting, demonstrates that this field scaling carries over to the resistivity components
\ba
\rho_{xx}(B) &=& \mu^{-1}\frac{\sqrt{1+\lp\frac{n_1-n_2}{n_1+n_2}\rp^2\lp\mu_H B\rp^2}}{\sqrt{n_1 n_2}}
\\ \rho_{xy}(B) &=& \frac{2B}{\lp n_1+n_2\rp}\lp\frac{\mu_H}{\mu}\rp
\ea
and also the magnetoresistance 
\ba
\label{binary MR}
\text{MR}(B) = \sqrt{1+\lp\frac{n_1-n_2}{n_1+n_2}\rp^2\lp\mu_H B\rp^2}-1\,\,\,.
\ea
If $\mu_H^{-1}(T)\propto\mu^{-1}(T)\propto\rho(T,B=0)\propto T$, then
\ba
\nn
\rho(B,T)=\frac{\sqrt{\lp\text{const.}\times T\rp^2 + \lp \frac{n_1-n_2}{n_1+n_2}\rp^2 \lp\text{const.}\times B\rp^2}}{\sqrt{n_1 n_2}}
\\ &\,&
\ea
indeed scales as a temperature/field quadrature combination; hence, this procedure provides one possible avenue for understanding the empirical formula seen in the aforementioned iron compounds.  Since the magnetic field only enters this model through its dimensionless combination $\lp\mu_H B\rp$ and the conductivity multiplicatively splits into two pieces $\sigma(B)=\sigma_0\times\tilde\sigma(\mu_H B)$, it automatically follows that the magnetoresistance admits single-parameter scaling MR$(B,T)=f(\mu_H B)=f(\# B/T)$ in this combination.  Note that in this special, equally distributed, discrete binary case, there exists a stricter single-parameter scaling in the combination of dimensionless field strength $\lp\mu_H B\rp$, {\it and} the "disorder strength" $\eta:=|(n_1-n_2)/(n_1+n_2)|$  in terms of the single parameter $x=\lp\eta\mu B\rp$ such that $MR(\mu,B,\eta)=\sqrt{1+x^2}-1$.

\begin{figure}[h!]
	\centering
	\subfloat[\label{sfig:5a}]{
		\includegraphics[width=.49\linewidth]{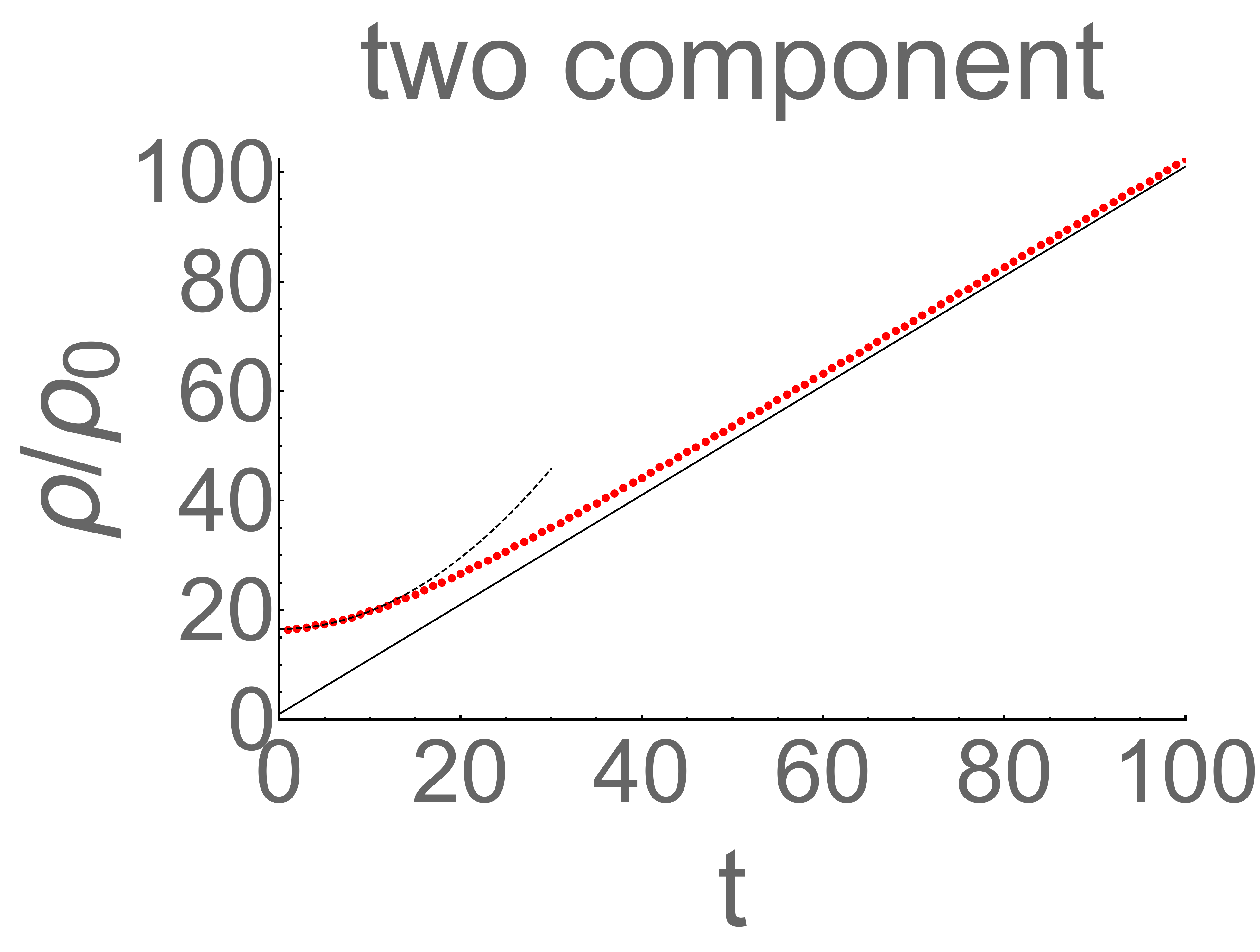}
	}
	\subfloat[\label{sfig:5b}]{
		\includegraphics[width=.49\linewidth]{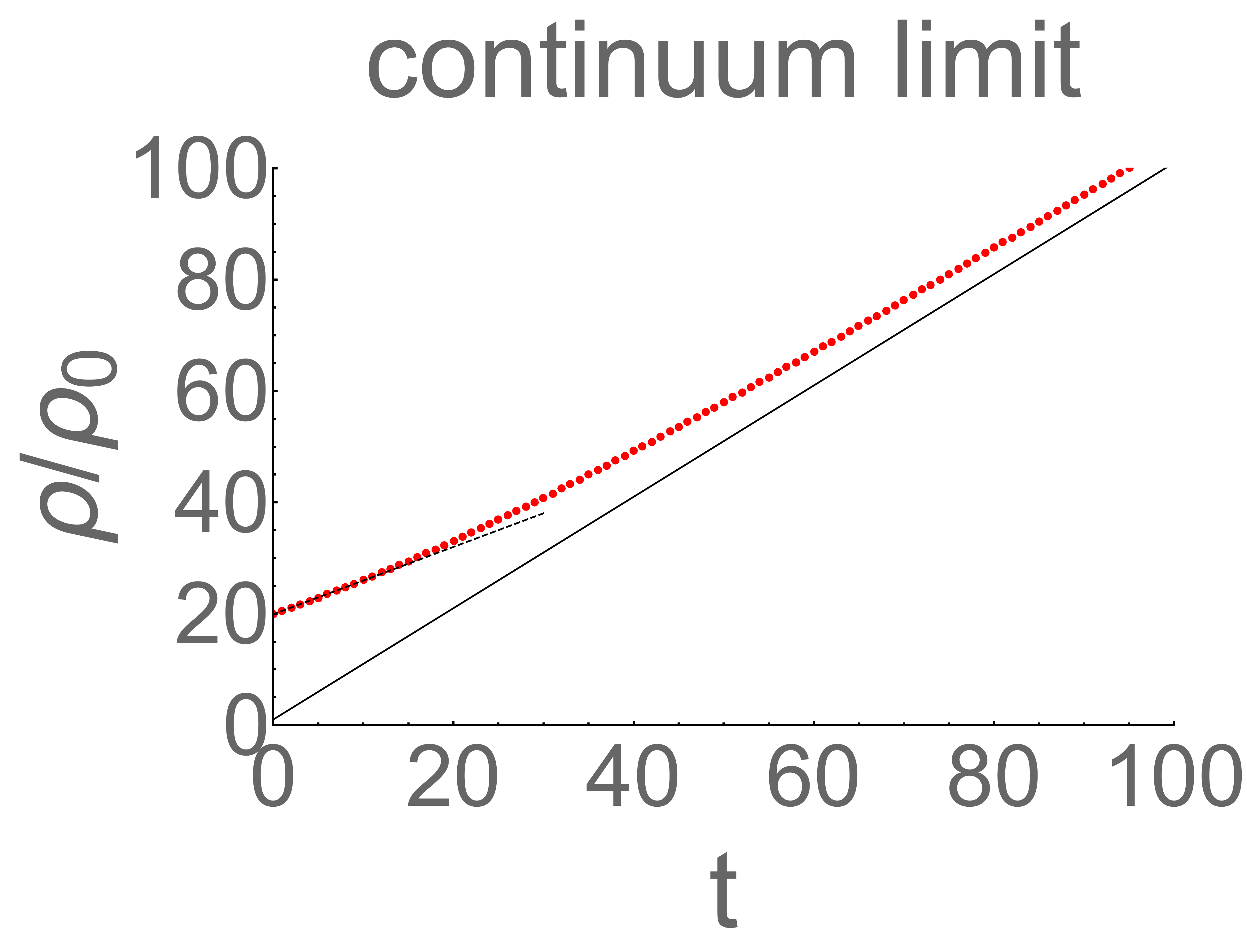}
	}

	\subfloat[\label{sfig:5c}]{
		\includegraphics[width=.32\linewidth]{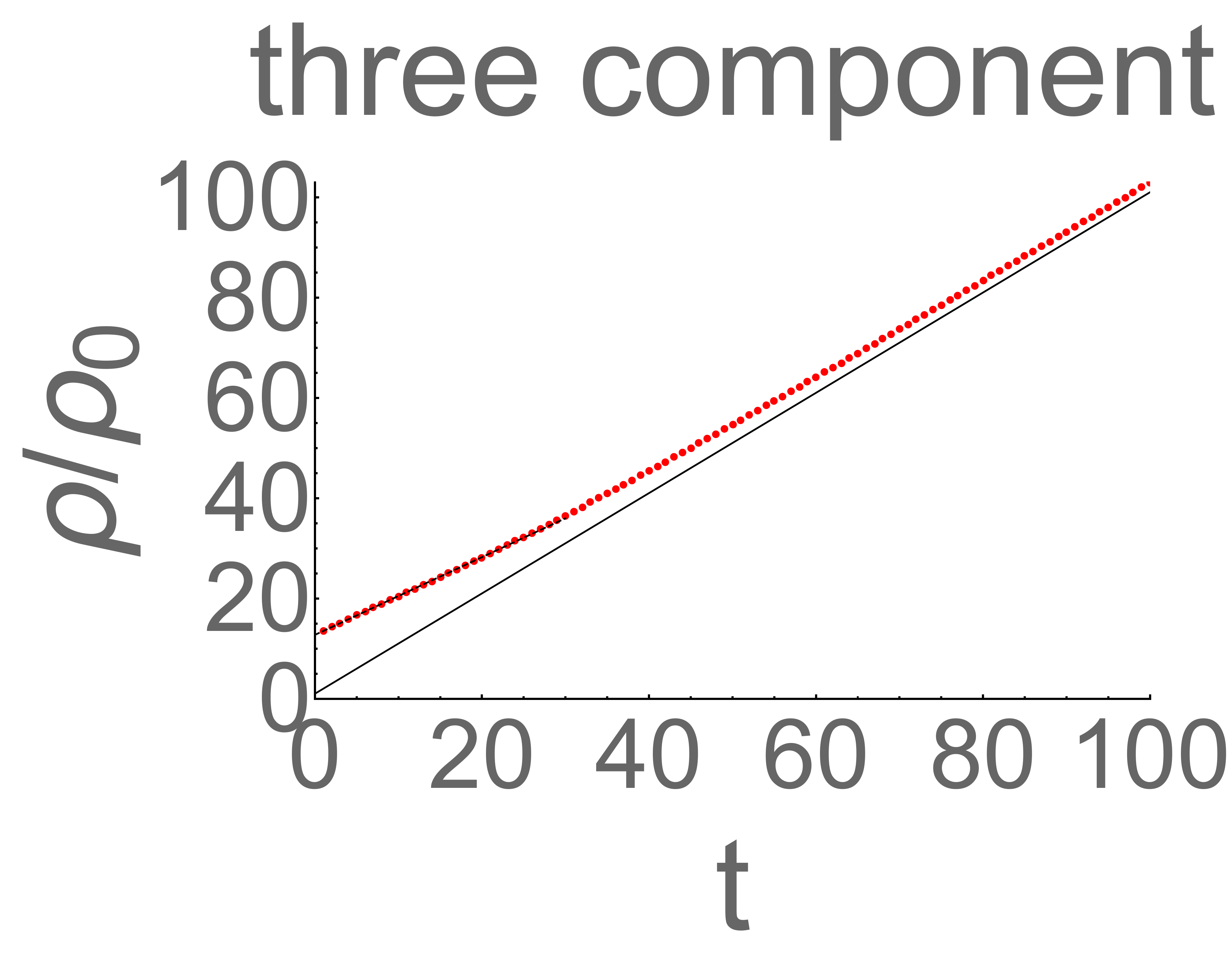}
	}
	\subfloat[\label{sfig:5d}]{
		\includegraphics[width=.32\linewidth]{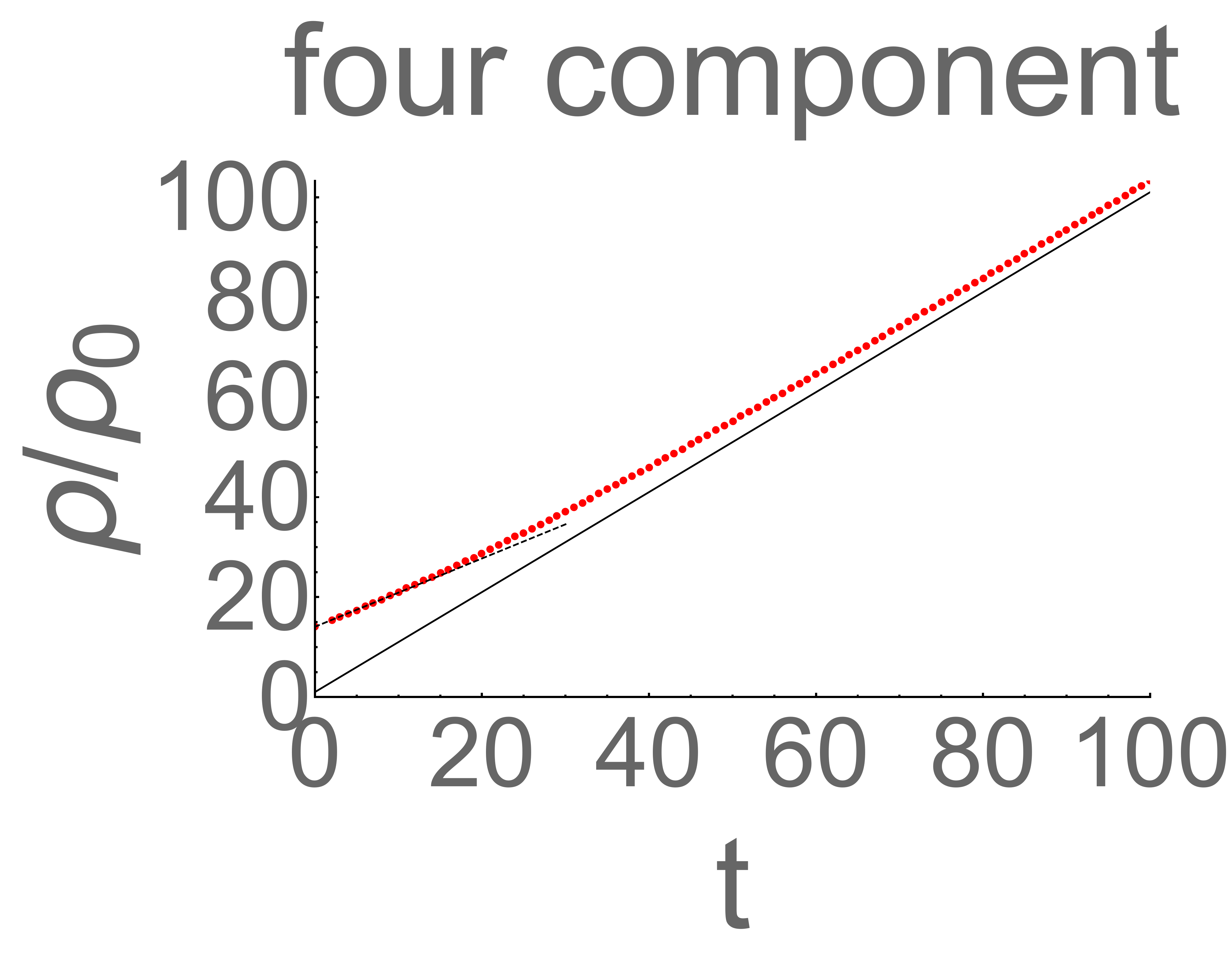}
	}
	\subfloat[\label{sfig:5e}]{
		\includegraphics[width=.32\linewidth]{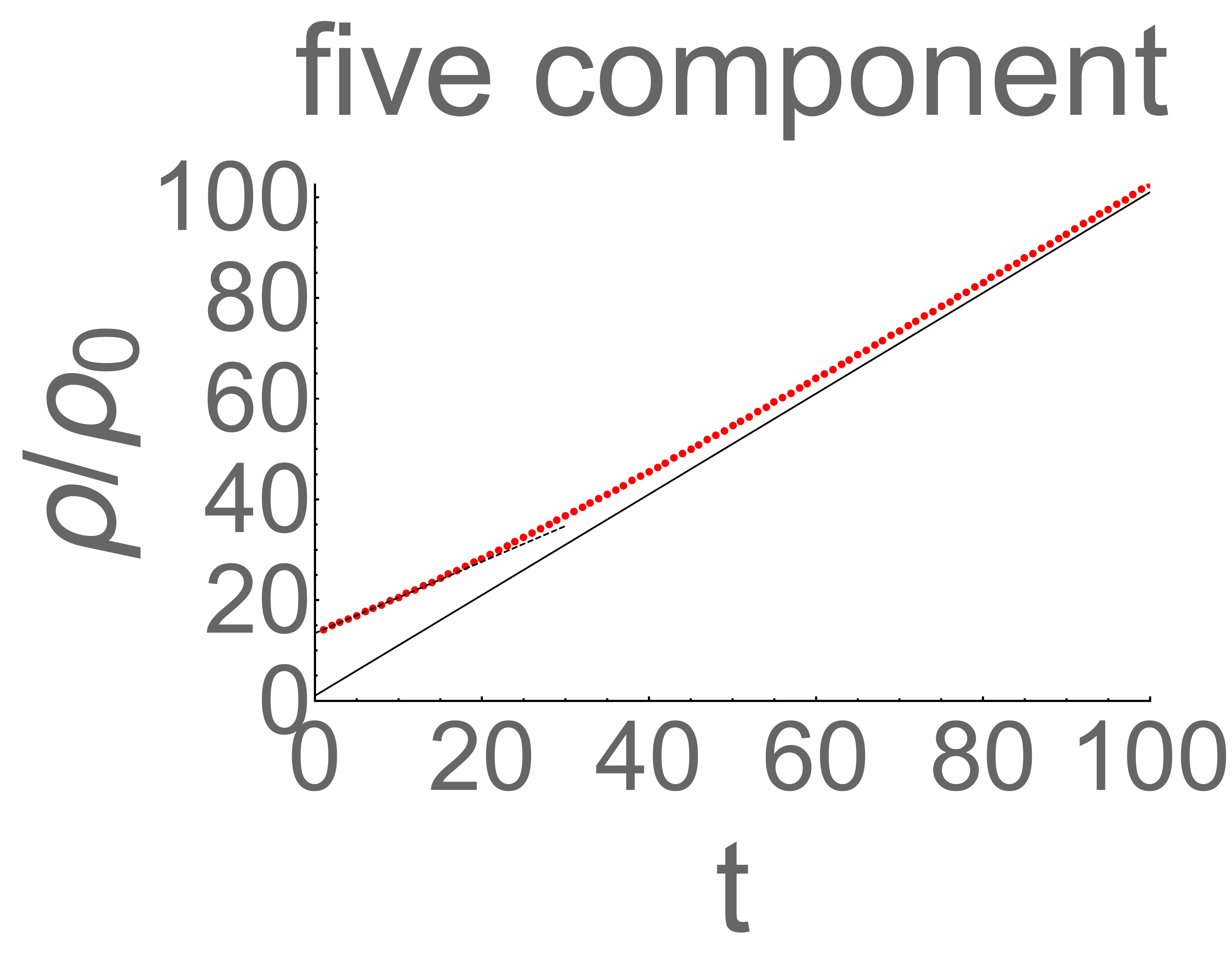}
	}
	\caption{{\bf (a)-(e)}  The scaled resistivity $\rho(T,B)/\rho_0$ for evenly-distributed, evenly-spaced discrete (a) 2 component, (b) continuum limit, (c) 3 component, (d) 4 component, and (e) 5 component metallic distributions in a strong magnetic field.  Dashed lines are (a) quadratic and (b)-(e) linear fits to the low $t$ resistivities,  while the black lines are the zero field curves for a $T-$linear resistivity modeled by $\rho(T) = \rho_0+\alpha T= \rho_0\lp 1+t\rp$.}
	\label{fig5}
\end{figure}

We now extend the same analysis to three, four, and five component, equally distributed metallic systems.  Of the discrete cases examined, only the binary (two-component) case has a non-linear low-temperature resistivity that is asymptotically quadratic in temperature (Fig. \ref{fig5}) as $T\to0$ despite the fact that each features a qualitatively quadratic-to-linear magnetoresistance at fixed temperature (and/or field scaling $\mu_H$).   It may be quite general of quadratic-to-linear MR curves, obtained from an effective medium approximation or otherwise, to produce low-temperature, high-field $T,B$-linear resistivities for systems with a $T$-linear zero-field resistivity.  That the low-temperature resistivity should generally be set by the zero-field resistivity in disorder-based models of the magnetoresistance was emphasized previously in \cite{Singleton18}.  Consequently, the simple square-root function currently appears as unique to the two-component inhomogeneous system.

We further bolster the generality of simultaneously T,B-linear regions being typical outputs of an effective medium approximation by examining the continuum limit of these equally spaced, evenly distributed discrete systems -- i.e. a system with a box probability distribution governing the carrier densities.  The coupled effective-medium equations become
\ba
\label{flat 1}
\int_{n_0-\Delta n_0}^{n_0+\Delta n_0} dn\, \frac{n^2-\lp n_{xx}^E\rp^2+\lp\mu B\rp^2\lp n-n_{xy}^E\rp^2}{\lp n+n_{xx}^E\rp^2+\lp\mu B\rp^2\lp n-n_{xy}^E\rp^2}
= 0
\\ \label{flat 2}
\int_{n_0-\Delta n_0}^{n_0+\Delta n_0} dn \,\frac{n-n_{xy}^E}{\lp n+n_{xx}^E\rp^2+\lp\mu B\rp^2\lp n-n_{xy}^E\rp^2}
=0
\ea
which can be rescaled $n,n_{xx,xy}^E\to (n/n_0),(n_{xx,xy}^E/n_0)$ to depend only on the relative disorder strength $\Delta$:
\ba
\label{sce 1}
\int_{1-\Delta}^{1+\Delta } dn\, \frac{n^2-\lp n_{xx}^E\rp^2+\lp\mu B\rp^2\lp n-n_{xy}^E\rp^2}{\lp n+n_{xx}^E\rp^2+\lp\mu B\rp^2\lp n-n_{xy}^E\rp^2}
&=& 0
\\ \label{sce 2}
\int_{1-\Delta}^{1+\Delta} dn \,\frac{n-n_{xy}^E}{\lp n+n_{xx}^E\rp^2+\lp\mu B\rp^2\lp n-n_{xy}^E\rp^2}
&=& 0\,\,\,.
\ea
These integrals can be performed exactly to obtain non-linear equations
\ba
\nn
0 &=& -n_{xx}^E \Phi_1(n_{xx}^E,n_{xy}^E,\mu B,\Delta)-2n_{xx}^E \Phi_2(n_{xx}^E,n_{xy}^E,\mu B,\Delta)
\\
&\,& + 2\Delta\lb\lp\mu B\rp^2+1\rb
\\ 0 &=&
\lp\mu B\rp \Phi_1(n_{xx}^E,n_{xy}^E,\mu B,\Delta)-\Phi_2(n_{xx}^E,n_{xy}^E,\mu B,\Delta)
\ea
with the functions $\Phi_1$ and $\Phi_2$ defined as
\ba
\Phi_1 =\qquad\qquad\qquad\qquad\qquad\qquad\qquad\qquad\qquad\qquad&\,&
\\ \nn
\ln\lp\frac{\lp n_{xx}^E\rp^2+2n_{xx}^E(1+\Delta)+\lp\mu B\rp^2\Delta_{\rm xy}^2+(1-\Delta)^2}{\lp n_{xx}^E\rp^2+2n_{xx}^E(1-\Delta)+\lp\mu B\rp^2\Delta_{\rm xy}^2+(1-\Delta)^2}\rp&\,&
\\ 
\Phi_2 =
\tan^{-1}\lp\frac{\sigma_E+\lp\mu B\rp^2\lp 1+\Delta-\sigma_H\rp+1+\Delta}{\lp\mu B\rp\lp n_{xx}^E+n_{xy}^E\rp}\rp\qquad&\,&
\\ \nn- \tan^{-1}\lp\frac{n_{xx}^E+\lp\mu B\rp^2\lp 1-\Delta-n_{xy}^E\rp+1-\Delta}{\lp\mu B\rp\lp n_{xx}^E+n_{xy}^E\rp}\rp\qquad&\,&.
\ea
For compactness, we have made the substitution $\Delta_{\rm xy}=n_{xy}^E-1-\Delta$ in the above formulas.  The resistivity obtained from this continuum limit scales linearly in temperature (Fig. \ref{sfig:5b}) at strong fields (regime of linear MR) and low temperatures as found in the discrete cases with more than two metallic components.

\begin{figure}[h!]
	\centering
	\subfloat[\label{sfig:6a}]{
		\includegraphics[width=.49\linewidth]{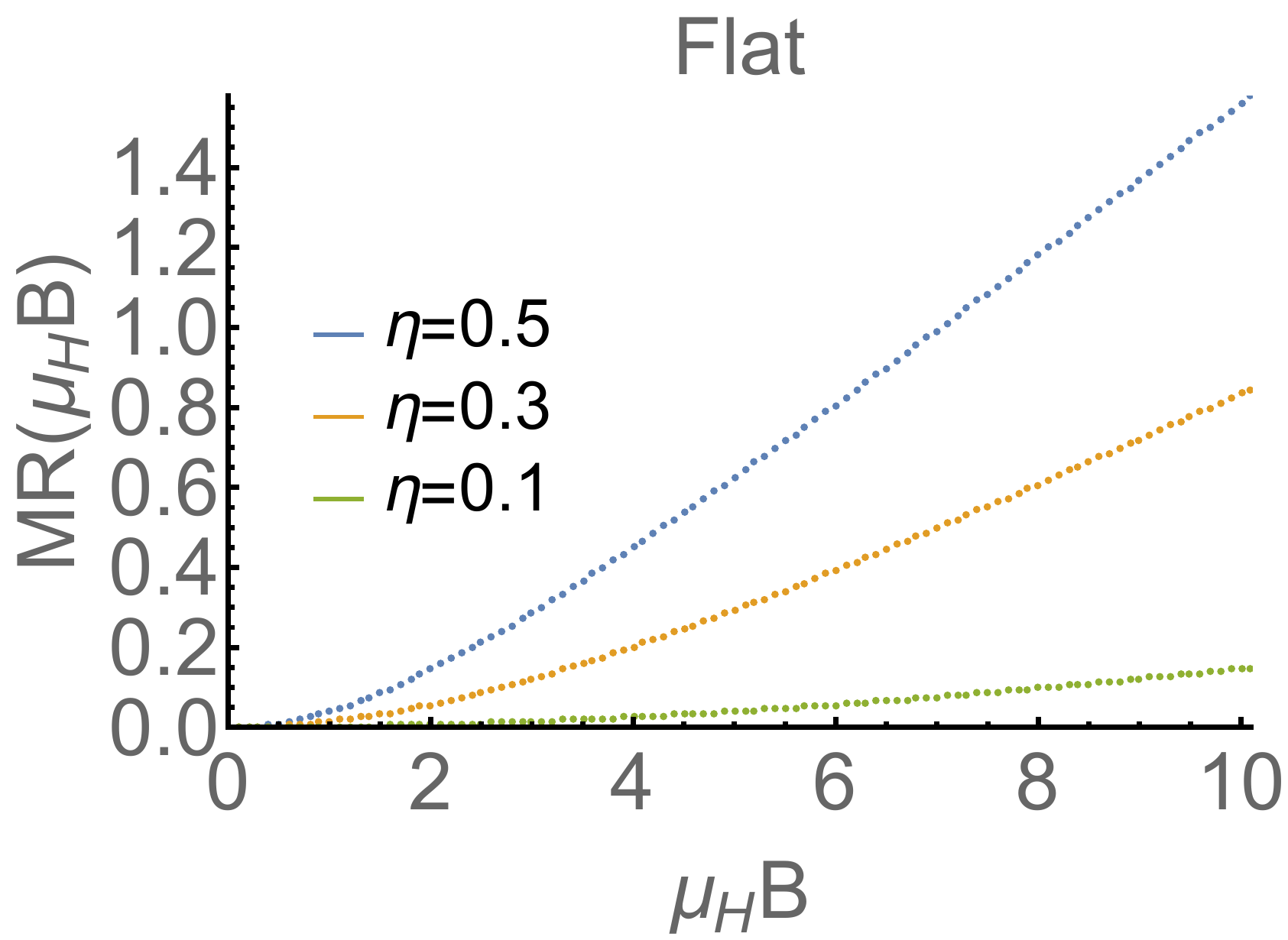}
	}
	\subfloat[\label{sfig:6b}]{
		\includegraphics[width=.49\linewidth]{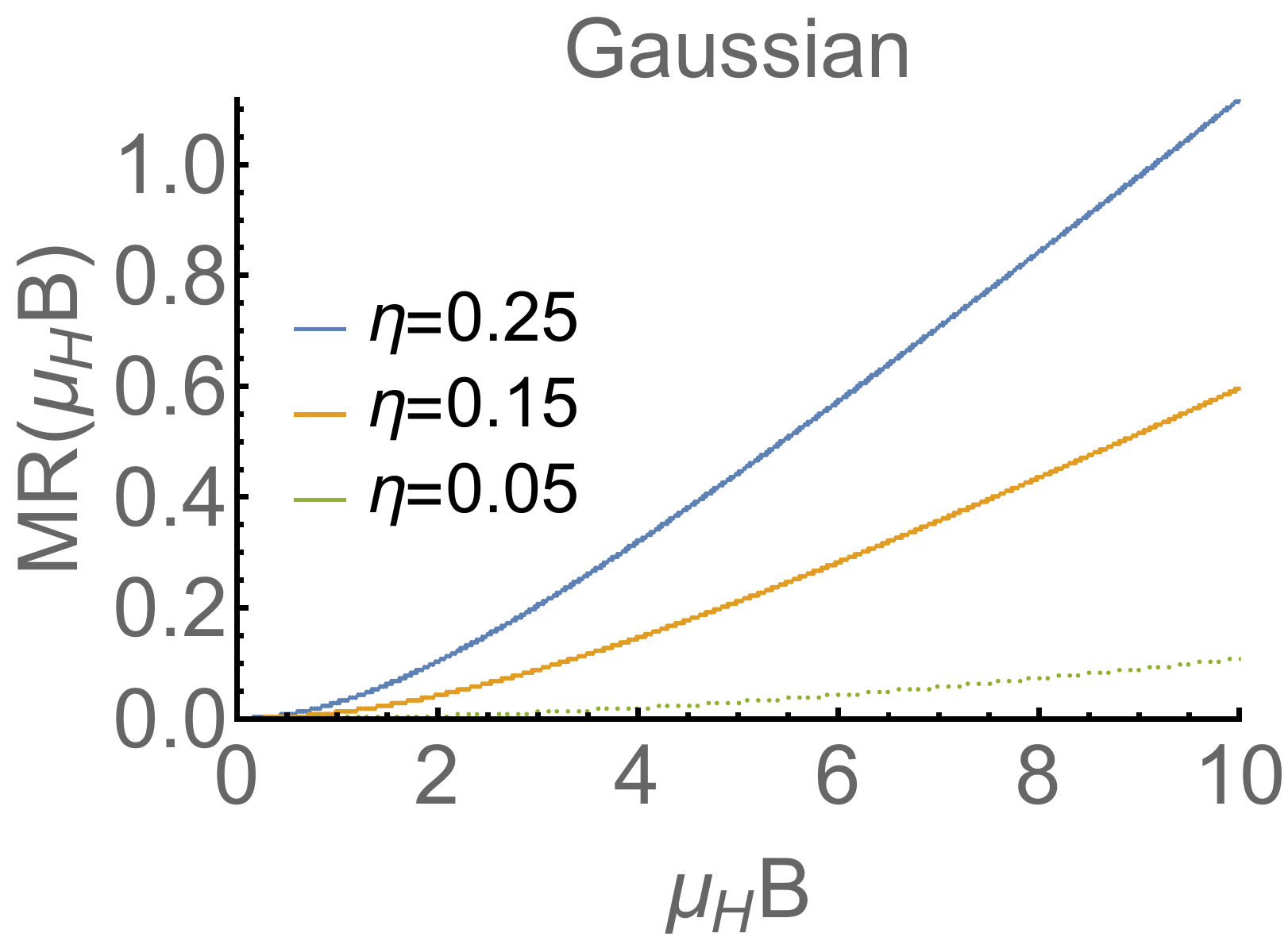}
	}

	\subfloat[\label{sfig:6c}]{
		\includegraphics[width=.49\linewidth]{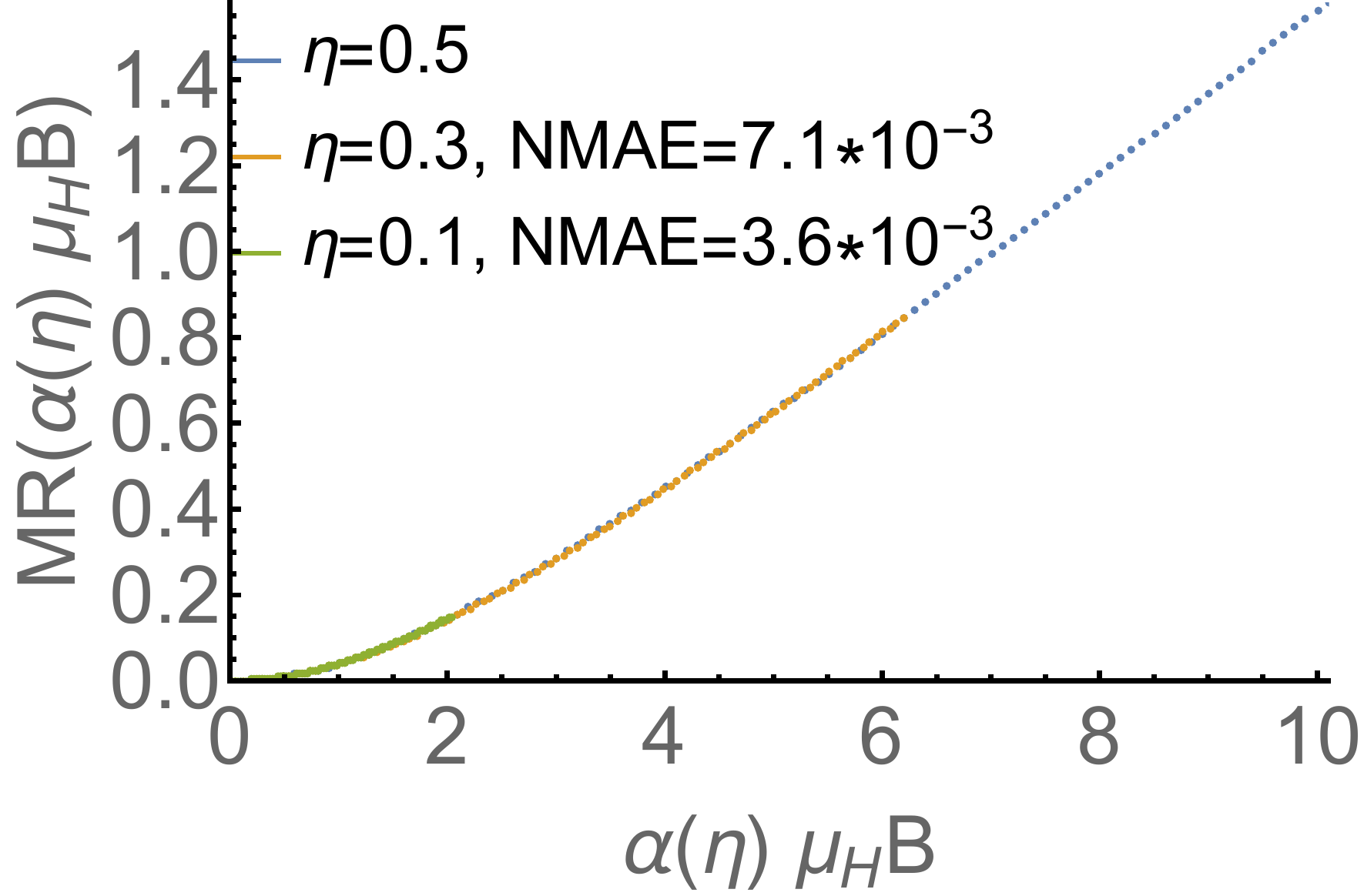}
	}
	\subfloat[\label{sfig:6d}]{
		\includegraphics[width=.49\linewidth]{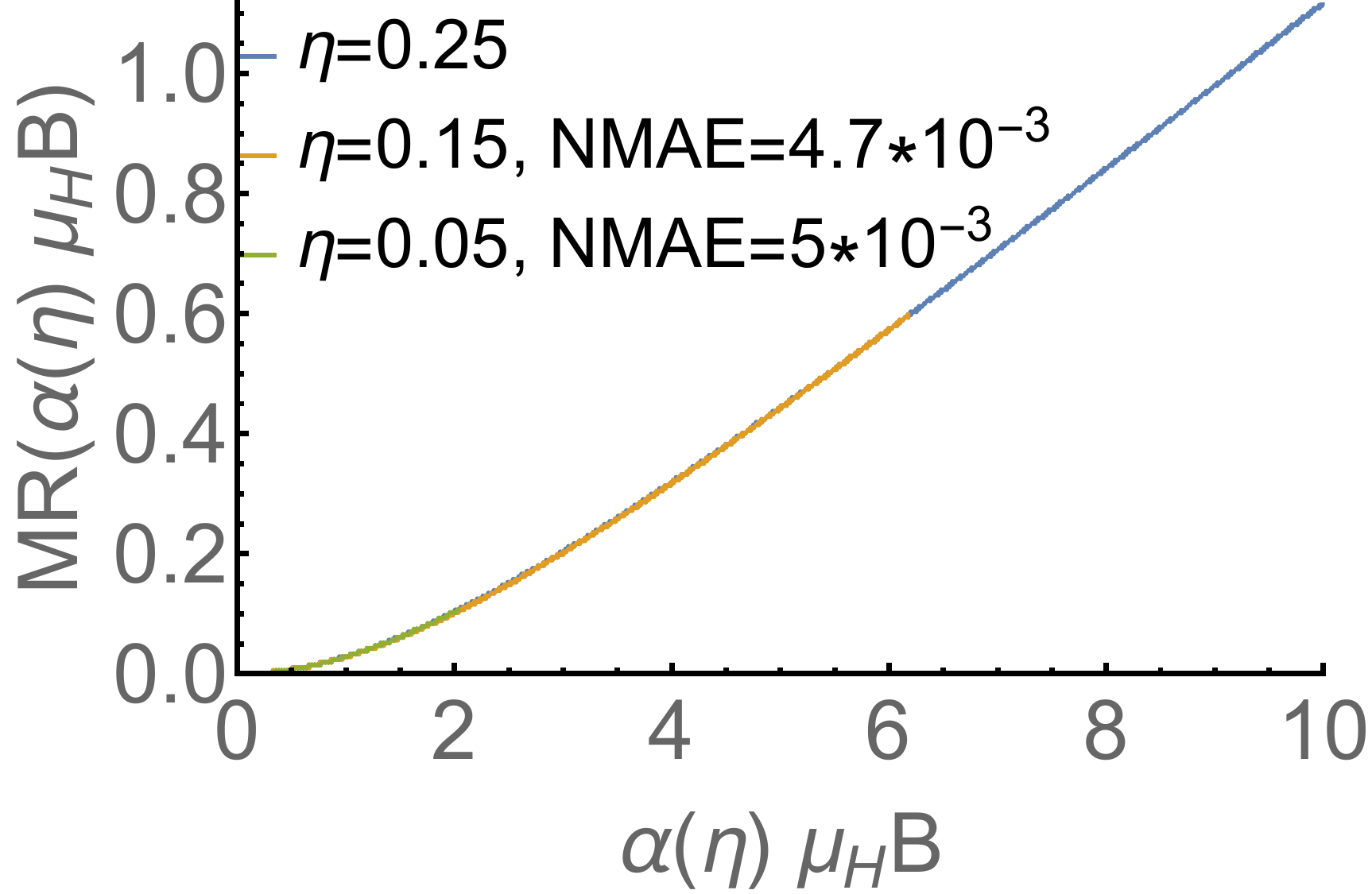}
	}
	\caption{{\bf (a),(b)}  The MR curves from the continuum limit of the discrete, equally distributed metallic systems (a) and including a Gaussian disorder profile (b).  {\bf (c),(d)}  The MR where the magnetic field strength is scaled by a disorder-dependent factor $\alpha(\eta)$ corresponding to the data plotted above.  The Normalized Mean Absolute Error (NMAE := [sum of $|$deviations of fit to data$|$]/[sum of $|$data$|$]) are obtained by fitting a scaled interpolating function of the less disordered curves to the calculated values at the strongest disorder strength $\eta$.  Discrepancy between left and right magnitudes of the curves is due to different effective definitions of $\eta$ in either the box or Gaussian case.}
	\label{fig6}
\end{figure}

The form of the magnetoresistance in the binary, discrete case \eqref{binary MR}, demonstrates that the disorder strength $\eta=(n_1-n_2)/(n_1+n_2)$ only enters coupled to $(\mu B)$.  This combination suggests that the MR curves obtained from the coupled effective medium equations \eqref{sce 1},\eqref{sce 2} might generically depend only on a single combination of some disorder-strength dependent function $\alpha(\eta)$ and the dimensionless magnetic field strength: MR$(\eta,\mu_H B)=f(\alpha(\eta) \mu_H B)$.  In continuum models, the disorder strength can be defined by the ratio of the variance to the mean of the underlying disorder distribution.  For finite discrete combinations, this additional disorder-dependent scaling appears unique to binary distributions.  In the continuum limit of the discrete cases and for the continuum Gaussian distribution, however, the single-parameter (disorder) dependence appears to re-emerge (Fig. \ref{fig6}).  Previously \cite{Ramakrishnan17}, Gaussian disorder distributions were shown to produce MR curves through either effective medium theory or averaging random resistor networks that produced equivalent curves when both horizontal and vertical axes were rescaled by disorder-dependent terms.  The scaling used to separately collapse the calculated Gaussian and box distribution MR curves, merely tuning a coefficient multiplying the horizontal scale $(\mu_HB)$, is less sensitive to experimental noise and makes comparison between data sets of unknown disorder strengths more direct.

It appears that the types of field and temperature dependence obtained from an effective medium theory involving equally distributed metallic components can be grouped into two categories (Fig. \ref{figemt}).  In the case of a random, equally distributed, binary combination, effective medium theory produces MR curves equivalent to the quadrature scaling seen in the doped iron compounds.  For all other equally distributed discrete or continuum cases (including an inhomogeneity profile governed by a Gaussian distribution), the functional form of the MR is such that the low-temperature, high-field resistivity remains $T$-linear toward $T=0$.  It is tempting to wonder if the binary case should be excluded as a pathology; however, accurately predicting the effective conductivity in binary metallic mixtures (suitably far from any percolative critical point) is a triumph of effective medium theory\cite{Landauer52,Kirkpatrick71}.  When the two metallic components differ in sign of charge carriers and are present in equal area fractions, the square root \eqref{binary MR} function is an exact result\cite{Guttal05} due to a duality theorem\cite{Mendelson75}.  For two-component systems with the same sign of carriers in each domain, another exact result\cite{Magier06} verifies the asymptotic coefficients of the square root function at high field, even if the area fractions are not exactly evenly distributed.  In light of the ability of metallic, two-component systems to produce the desired quadrature temperature/field resistivity scaling seen near the quantum critical point of the aforementioned iron compounds,  irrespective of the microscopic dynamics be they SYK or other non-quasiparticle descriptions, it may be worth seriously considering how such a seemingly specific description arises from critical fluctuations between ordered/disordered metallic phases.

\section{Application of effective medium theory to LSCO}

The prominent features reported within the experimental measurement of the LSCO data are a quadratic-to-linear unsaturating MR and the existence of a high field ($B$=50-80T), low temperature region where the resistivity displays a simultaneous $T,B$-linear change in resistivity.  As these appear to be generic features of effective medium models with more than two metallic constituents (Fig. \ref{fig5}), we examine how the assumptions within effective medium theory are met by the LSCO samples in question.  Naturally, this first requires demonstrating the existence of inhomogeneity, i.e. charge density variations, at the dopings where the MR data\cite{Boebinger18} were obtained.  Scanning tunneling microscopy (STM) at low temperatures \cite{Kato05} near optimally doped LSCO observed variations in the $T=4.2$K local spectral gap and $^{63}$Cu NQR measurements \cite{Singer02} observe variations in the local hole density at high temperatures $100$K$<T<300$K across a wide doping range.  In a cuprate with similar inhomogeneity, the local spectral gap was concretely found to track the local doping level at low temperature in Bi2201\cite{Hoffman12}, and so we assume this holds in LSCO as well.  The two measurements both find a length scale of 5-10 nm associated with the local inhomogeneities and infer, at low temperatures, an effectively temperature-independent local doping variation of $p_{\rm loc}=.12-.18$ near $p=.16$, which corresponds to a sizable disorder strength $\eta\approx.125$.  The mean free path in optimally doped LSCO is estimated through the standard resistivity formula\cite{Boebinger96} for a layered 2D, free electron system
\ba
l = \frac{h d}{\rho k_F e^2},
\ea
where $d\approx0.64$nm is the interlayer distance, $\rho$ the in-plane resistivity, and $k_F$ is a typical Fermi wavevector.  Across the measured dopings, $k_F$ doesn't change drastically\cite{Yoshida07}; in the nodal regions $k_F\approx5.3$nm$^{-1}$ and near the antinodal regions\cite{Hussey11} $k_F\approx7$nm$^{-1}$.  Using the experimental fit to $\rho(T)$ and the typical (anti)nodal wavevectors as bounds, we find that $l\approx1.3-1.7$nm at $T=180$K, $l\approx4-6$nm at $T=50K$, and that the mean free path equals roughly the maximal domain size $l\approx 10$nm at $T\approx 22-29$K.  At high temperatures, the mean free path $l$ is smaller than the typical size of density variations, which we take as justification to apply the {\it classical} conductivity averaging procedure described in effective medium theory.  Perhaps comfortingly, other macroscopic properties are known to vary locally in the BSCCO family of cuprates, such as the growing evidence in the past decade for nanoscale Fermi surface variations\cite{Wise09,Hoffman18,Alldredge13}.

\begin{figure}[h!]
	\centering
	\subfloat[\label{sfig:7a}]{
		\includegraphics[width=\linewidth]{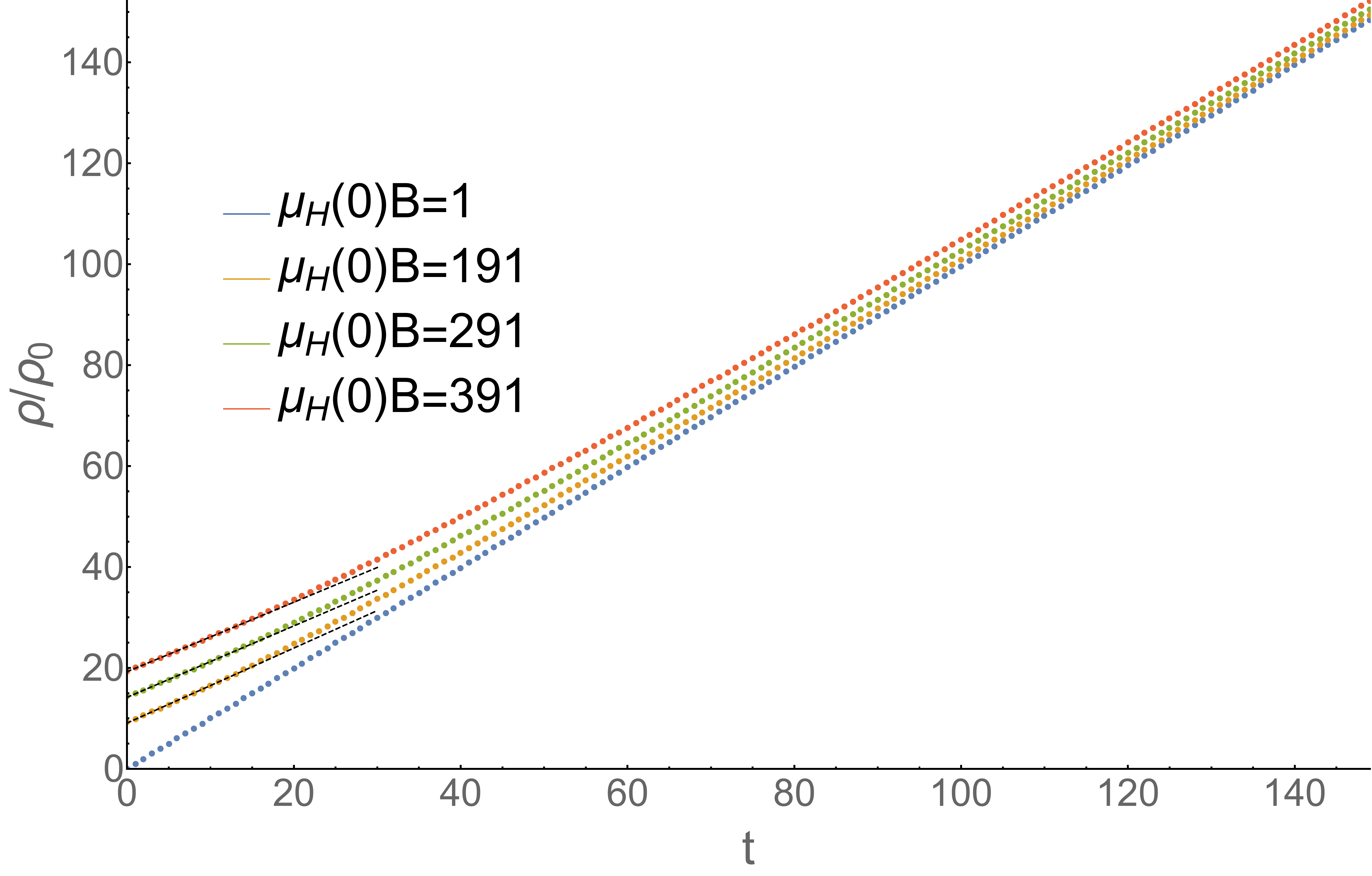}
	}

	\subfloat[\label{sfig:7b}]{
		\includegraphics[width=\linewidth]{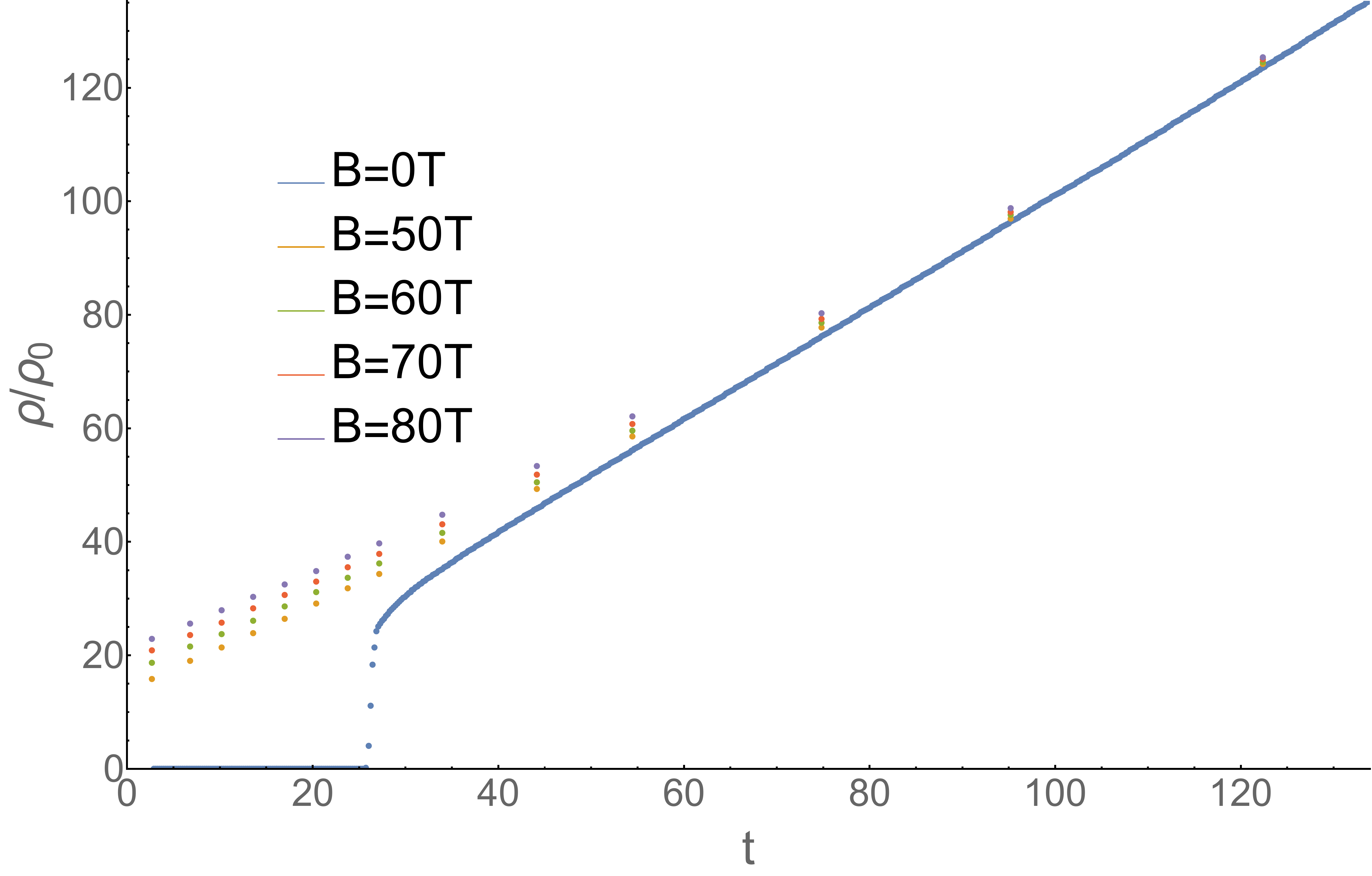}
	}
	\caption{{\bf (a)}  The dimensionless resistivity $\rho(t,\mu_H(t)B)/\rho_0$ as a function of dimensionless temperature $t$ and field strength $\mu_H(t)B$ where $\mu^{-1}_H(t)\propto\rho(t,B=0)/\rho_0=(1+t)$ calculated using a Gaussian disorder distribution.  Black dashed lines are linear fits to the low-temperature data.  {\bf (b)}  The low-temperature resistivity from experiment, rescaled by $\rho_0\approx 1.5\mu\Omega$cm and using $t(T)=(\alpha T/\rho_0)\approx [0.68$K$^{-1}] T.$}
	\label{fig7}
\end{figure}

\begin{figure}
	\includegraphics[width=\linewidth]{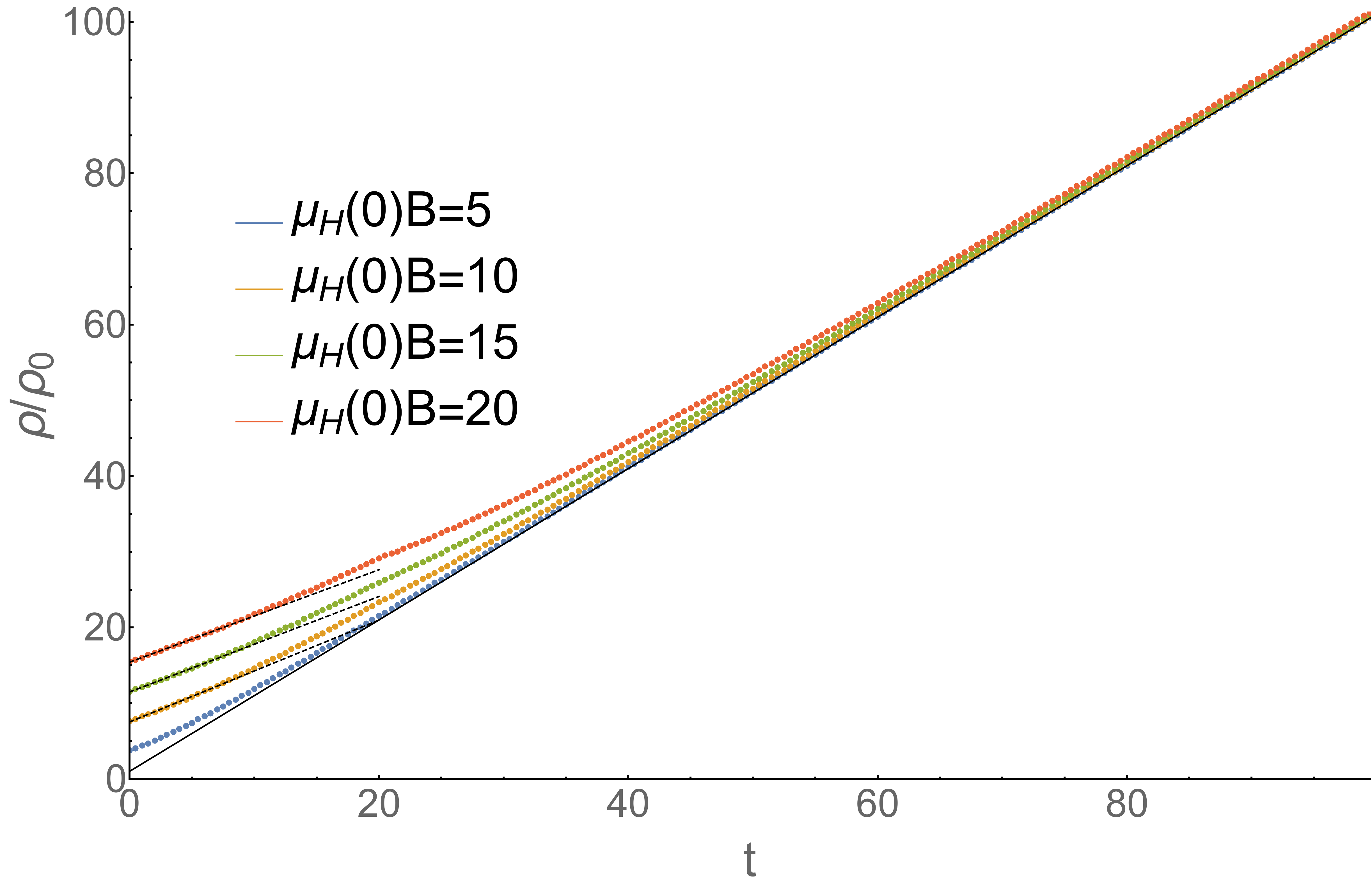}
	\caption{The dimensionless resistivity $\rho(t,\mu_H(t)B)/\rho_0$ where the temperature scaling is a simple piecewise combination of $\mu_H^{-1}(t)\propto(1+t)$ for $t\le20$ and $\mu_H^{-1}(t)$ proportional to the nonlinear fit $\mu_H^{-1}(t)\propto1+at+bt^{5/3}$ (Fig. \ref{sfig:2b}) for $t\ge20$.  The kink in the topmost $\mu_H(0)B=20$ curve is due this simple piecewise model.  At higher $\mu_H(0)B$ values, the resistivity will eventually upturn at low temperature.  The variation in the magnitude of  $\mu_H(0)B$ arises from the use of a box disorder distribution rather than a Gaussian as in Fig. \ref{sfig:7a}.}
	\label{fig8}
\end{figure}

The experimental fit to the $T$-linear zero-field resistivity of optimally-doped LSCO $\rho(T)\approx \rho_0+\alpha T$, with $\alpha\approx1.02\mu\Omega$cm/K and $\rho_0\approx1.54\mu\Omega$cm, allows for a dimensionless temperature $t=\alpha T/\rho_0$ ($\alpha/\rho_0\approx 0.68$K$^{-1}$ in the measured sample) and scaled resistivity $\rho/\rho_0$ to be constructed.  If we take the inverse temperature scale that couples to the magnetic field to also scale linearly in temperature, $\mu_H^{-1}(T)\propto\rho(T,B=0)$, then the resistivity $\rho(T,B)$ curves obtained through an effective medium calculation are qualitatively similar to the experimental data measured at strong magnetic fields and low temperatures (Fig. \ref{fig7}).  As found earlier (Fig. (\ref{sfig:2b}), however, the high-temperature ($50<T<180$) LSCO MR data indicates that the field scaling  $\mu^{-1}_H(T)$ is non-linear.  So long as this behavior transitions to a linear form at the lowest temperatures, this scaling can continue to hold (Fig. \ref{fig8}).  The effect of the high temperature curvature in $\mu_H^{-1}(T)$ can be seen by comparing how rapidly the curves at different field strength $\mu_H(0)B$ approach one another (Fig. \ref{fig8} vs. Fig. \ref{sfig:7a}).

As mentioned in the previous discussion on MR curve collapse, a rapidly changing zero-field resistivity around $T\sim20$K -- in contrast to a rapidly changing $\mu_H(T)$ -- offers one possible way of reconciling the lack of curve collapse at low temperature $T\le20$K (Fig. \ref{fig3}). Yet, the resistivity curves in optimally-doped LSCO (Fig. \ref{fig7}) appear smooth around $T=20$K.  Presumably, the single-parameter scaling seen at higher temperatures must end below $T\sim20$K.  In the context of a nanoscale inhomogeneity description for the MR in LSCO, this is to be expected.  The calculated mean free path $l$ is on the order of the largest density domains near $T\approx 22-29$K, below which -- for $T=15$K, 10K, 4K -- curve collapse appears to be broken and the use of a local, classical conductivity averaging procedure becomes suspect.  Near $T\sim 25$K, the field slope of the resistivity, $\beta=\partial\rho/\partial B$, was found to saturate to a temperature-independent value.  Similar abrupt changes in the low-temperature MR are seen in a disordered MnAs-GaAs composite\cite{Bennett10}, where the single-parameter scaling, inferred from $\partial$(MR)$/\partial T\propto\mu_H(T)$ in the B-linear MR regime, breaks down once the mean free path increases above the average spacing between the MnAs nanoparticles.  In this compound, the mobility $\mu_H(T)=\mu(T)$ was measured separately and found to be proportional to the linear MR field slope from $T=50-300$K.  The experimental findings on MnAs-GaAs were understood within the context of a random resistor network\cite{Parish03,Parish05} whose conclusions in terms of single parameter scaling with $\mu_H(T)$ are identical to the EMT description\cite{Ramakrishnan17} within its regime of validity.

An EMT description of the MR in LSCO doesn't clarify why single-parameter scaling should only appear near the doping $p=.19$.  As far as the effective medium theory is concerned, the mean doping level is not particularly special; the only consideration is whether or not the sample contains inhomogeneity that varies on a sufficiently large scale so that it can be modeled as a random mixture of conducting patches.  From the $^{63}$Cu experiment \cite{Singer02}, the variation in local hole concentration is present in the lightly over/underdoped samples and increases on the over-doped side (measured at $p=0.2,0.25)$. Nothing appears unique about $p=.19$ in terms of the disorder profile.  The worsening curve collapse in the underdoped samples might follow from new mechanisms associated with the pseudogap order influencing the MR.

\begin{figure}[h!]
	\includegraphics[width=\linewidth]{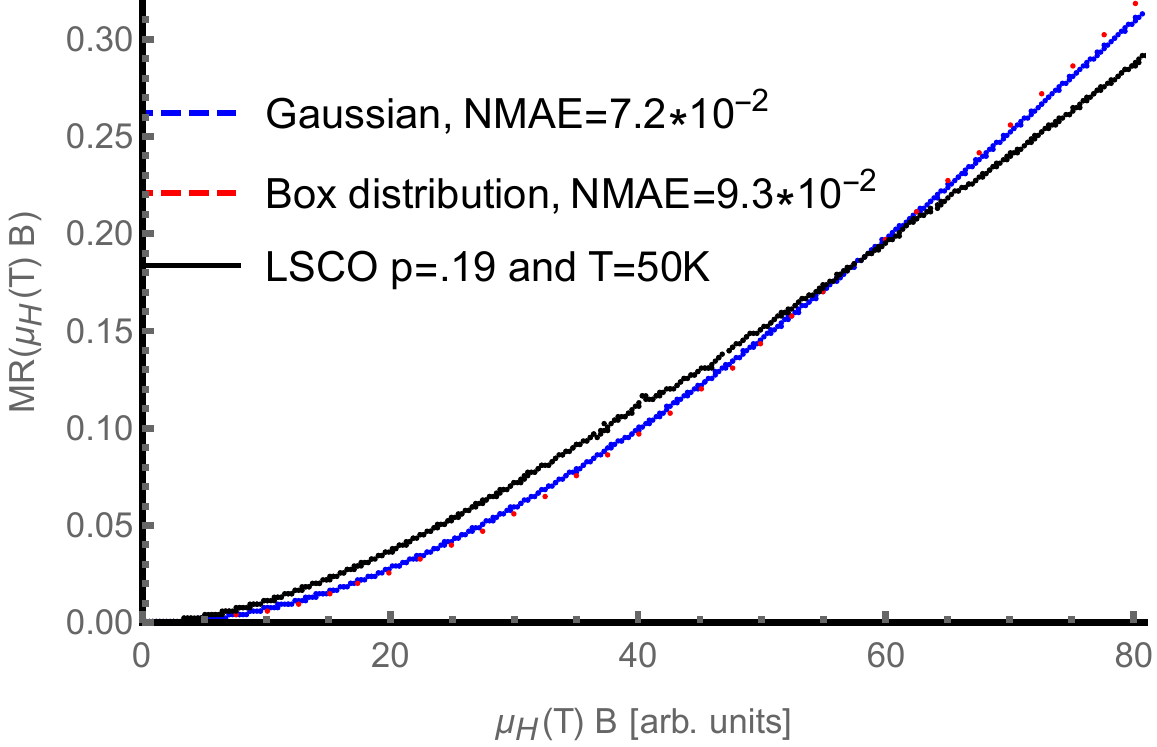}
	\caption{Comparison of the computed MR curves from effective medium theory, using either a Gaussian or box distribution, to the MR curve measured in LSCO doped to $p=.19$ at $T=50$K from $B=0$T-80T.}	
	\label{fig9}
\end{figure}

While the EMT with a continuum inhomogeneity profile qualitatively reproduces a quadratic-to-linear MR, single-parameter scaling in the MR, low-temperature high-field $T-$linear resistivity and accurately predicts when the curve collapse breaks down, the MR curves are not exactly identical to those obtained from the optimally-doped LSCO sample (Fig. \ref{fig9}) even when compared above $T_c$.  If the LSCO MR were perfectly predicted by the effective medium theory, the ability to scale out disorder (Fig. \ref{fig6}) should result in calculated and experimental MR curves that can be scaled onto each other by only tuning $(\mu B)$.  This appears not to be the case since the calculated and experimentally measured MR curves are more distinct than MR curves in the underdoped samples were (Fig. \ref{fig4}).  One possible source of error found in the nematic iron compound\cite{Hussey19} is that samples with a smaller residual resistivity see an enhanced quadratic-in-B MR from standard Fermi surface orbits associated with Kohler's rule.  In the optimally-doped sample ($p=0.19$), the residual resistivity is lower than in any of the other measured samples at other dopings (see Supplementary Materials\cite{Boebinger18}).  As a result, more traditional orbital mechanisms may be altering the low-field form of the MR curve as well as any of the following mechanisms not included: correlations between the density regions, the Fermi surface structure beyond the isotropic classical steady state equation used to model the DC current response, field-dependent scattering, and mobility variations associafted with the change in local doping level.  

\section{Summary}

We have demonstrated that the magnetoresistance (MR) in optimally doped LSCO above $T_C$ is governed by single-parameter scaling MR$(B,T)=f(\mu_H(T)B)$ for some function $f(x)$ distinct from the quadrature combination of field and temperature seen in the resistivity of iron materials near their quantum critical points\cite{Shekhter16,Hussey19,Nakajima19}.  We stress that this is a model-independent, empirical fact of the data set\cite{Boebinger18}.  The temperature scaling $\mu^{-1}_H(T)$ inferred from the curve fitting at high (and low) temperatures is not directly related to the zero-field resistivity $\rho(T)\sim T$ or the high temperature Hall angle $\cot_H(T)\sim T^2$, though it may be in a crossover regime between the two.

While the $T$-linear resistivity scaling in LSCO may have quantum critical origins, the small mean free path of this bad metal reasonably allows us to analyze its magnetoresistance within a classical model.  We find that an effective medium approximation, which takes experimental LSCO zero-field transport and its inhomogeneity profile as inputs, is capable of producing MR curves that qualitatively replicate the behavior of LSCO above $T\sim20$K.  At temperatures below $T\sim20$K, the mean free path is no longer smaller than the size of density variations and the breakdown of the classical conductivity averaging procedure occurs simultaneously with the lack of single-parameter scaling in the MR and the resistivity field slope $\beta=\partial\rho/\partial B$ saturating to a temperature independent value.  Interestingly, an inhomogeneous binary combination of metallic components appears to be unique in producing the previously mentioned quadrature combination of temperature and field dependence of the resistivity if both metallic components are characterized by a $T$-linear zero-field resistivity.  This case may be relevant to the iron-pnictides, but a broader-range of disorder seems to be necessary to capture the behaviour in the cuprates.  Hence, this work seems to hint at two operative mechanisms for the quadrature magnetoresistance observed in the pnictides and the unsaturating independent $T$ and $B$-linear resistivities in the cuprates.

\section*{Acknowledgments}
The authors would like to thank Navneeth Ramakrishnan, Arkady Shekhter, and Luke Yeo for helpful comments.  P.W.P. would like to thank NSF DMR-1461952 for partial funding of this project.

\bibliographystyle{apsrev4-1}

\end{document}